\documentclass[sigconf]{acmart}
\AtBeginDocument{%
  }

\setcopyright{acmlicensed}
\copyrightyear{2018}
\acmYear{2018}
\acmDOI{XXXXXXX.XXXXXXX}
\acmConference[Conference acronym 'XX]{Make sure to enter the correct
  conference title from your rights confirmation email}{
  2018}{Woodstock, NY}
\acmISBN{978-1-4503-XXXX-X/2018/06}

\usepackage[utf8]{inputenc} 
\usepackage{makecell}

\usepackage{booktabs}
\usepackage{graphicx}
\usepackage{bm}  
\usepackage{hyperref}
\usepackage{multirow}
\usepackage{bm}
\usepackage{enumitem} 
\usepackage{colortbl}
\usepackage{xcolor,colortbl}
\usepackage{subcaption}

\usepackage{stfloats}
\usepackage{xurl} 
\usepackage{float}

\begin{document}

\title{Beyond Static Snapshots: Dynamic Modeling and Forecasting of Group-Level Value Evolution with Large Language Models}

\settopmatter{authorsperrow=4}

\author{Qiankun Pi}
\affiliation{%
  \institution{Wuhan University}
  \city{Wuhan}
  \state{Hubei}
  \country{China}
}
\email{piqiankun@whu.edu.cn}

\author{Guixin Su}
\affiliation{%
  \institution{Wuhan University}
  \city{Wuhan}
  \state{Hubei}
  \country{China}
}
\email{cometsue@whu.edu.cn}

\author{Jinliang Li}
\affiliation{%
  \institution{Wuhan University}
  \city{Wuhan}
  \state{Hubei}
  \country{China}
}
\email{lijinliang@whu.edu.cn}

\author{Mayi Xu}
\affiliation{%
  \institution{Wuhan University}
  \city{Wuhan}
  \state{Hubei}
  \country{China}
}
\email{xumayi@whu.edu.cn}

\author{Xin Miao}
\affiliation{%
  \institution{Wuhan University}
  \city{Wuhan}
  \state{Hubei}
  \country{China}
}

\email{miaoxin@whu.edu.cn}

\author{Jiawei Jiang}
\affiliation{%
  \institution{Wuhan University}
  \city{Wuhan}
  \state{Hubei}
  \country{China}
}

\email{jiawei.jiang@whu.edu.cn}

\author{Ming Zhong}
\affiliation{%
  \institution{Wuhan University}
  \city{Wuhan}
  \state{Hubei}
  \country{China}
}

\email{clock@whu.edu.cn}

\author{Tieyun Qian}
\affiliation{%
  \institution{Wuhan University}
  \city{Wuhan}
  \state{Hubei}
  \country{China}
}
\affiliation{%
  \institution{Zhongguancun Academy}
  \city{Beijing}
  \country{China}
}
\email{qty@whu.edu.cn}


\begin{abstract}
Social simulation is critical for mining complex social dynamics and supporting data-driven decision making. LLM-based methods have emerged as powerful tools for this task by leveraging human-like social questionnaire responses to model group behaviors. Existing LLM-based approaches predominantly focus on group-level values at discrete time points, treating them as static snapshots rather than dynamic processes. However, \textit{group-level values are not fixed but shaped by long-term social changes}. Modeling their dynamics is thus crucial for accurate social evolution prediction—a key challenge in both data mining and social science. This problem remains underexplored due to limited longitudinal data, group heterogeneity, and intricate historical event impacts.

To bridge this gap, we propose a novel framework for group-level dynamic social simulation by integrating historical value trajectories into LLM-based human response modeling. We select China and the U.S. as representative contexts, conducting stratified simulations across four core sociodemographic dimensions (gender, age, education, income). Using the World Values Survey, we \textbf{construct a multi-wave, group-level longitudinal dataset} to capture historical value evolution, and then \textbf{propose the first event-based prediction method} for this task, unifying social events, current value states, and group attributes into a single framework. Evaluations across five LLM families show substantial gains: a maximum 30.88\% improvement on seen questions and 33.97\% on unseen questions over the Vanilla baseline. We further find notable cross-group heterogeneity: U.S. groups are more volatile than Chinese groups, and younger groups in both countries are more sensitive to external changes. These findings advance LLM-based social simulation and provide new insights for social scientists to understand and predict social value changes.

\end{abstract}




\maketitle

\section{Introduction}
Social simulation has emerged as a fundamental methodology in sociology, providing a computational framework to probe complex behaviors and validate theoretical hypotheses~\cite{richiardi2006common,edelmann2020computational,harrison2007simulation}. Traditional simulations rely primarily on manual surveys, which inherently limit their scalability and behavioral granularity~\cite{davis1993general}. To mitigate this constraint, Large Language Models (LLMs) have been widely deployed to simulate diverse populations, enabling researchers to replicate large-scale surveys via prompting LLMs to respond questionnaires, thereby capturing group value distributions~\cite{park2024diminished,xu2024wizardlm}.

To further improve the accuracy of LLM-based social simulations, researchers typically adopt persona-based frameworks with multiple demographic dimensions, such as country and gender. By incorporating specific demographic attributes directly into the input context, these frameworks assign distinct social identities and cultural backgrounds to LLMs, enabling models to effectively capture the unique perspectives and value orientations to diverse social groups~\cite{cao2025specializing,liu2025towards,hwang2023aligning}. However, existing LLM-based social simulations only characterize group-level values at discrete moments, overlooking the dynamic and evolving nature of social values.

\begin{figure}[ht]
    \centering
    \includegraphics[width=1\columnwidth]{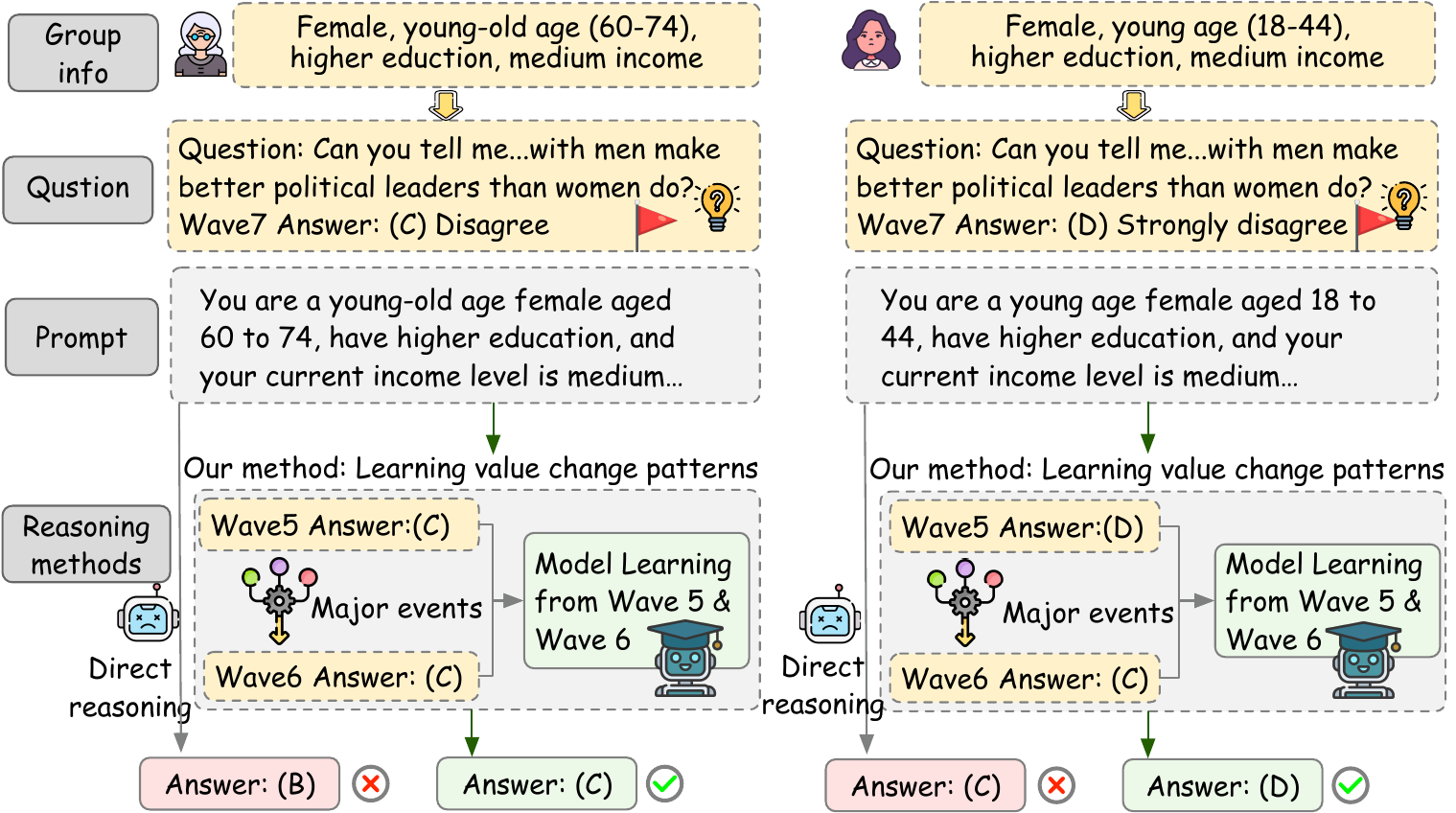}
    \caption{A example of two social groups in the United States from the World Values Survey (WVS), reflecting the dynamic evolution of social values.}
    \label{fig:intro}
\end{figure}

Consistent with sociological theories, human value beliefs are products of continuously evolving processes rather than static entities~\cite{hills2019historical,durkheim2023rules,mesoudi2008multiple,mesoudi2006towards,cavalli1981cultural}. As shown in Figure~\ref{fig:intro},  young-old females' responses to a specific question remain stable across the 5th–7th survey waves, whereas young females' exhibit noticeable changes. This contrast demonstrates that group-level values can either change over time or remain stable, which highlights the necessity of incorporating temporal dynamics into social simulation. Accordingly, effective LLM-based social simulation requires the ability to identify whether value change occurs and to model its evolutionary trajectories, a capability that is currently lacking in existing approaches.

To fulfill the requirement of dynamic social simulation, we enhance LLM-based social value survey simulations by integrating historical value trajectories into human response modeling, accounting for the \textbf{dynamic evolution of societal values over time}. Specifically, we aggregate populations into distinct demographic groups based on four core sociodemographic dimensions: gender, age, highest education level, and income level. This design overcomes the limitations of cross-sectional data, where tracking the same individuals over time from surveys is infeasible due to privacy constraints. Our group-level perspective enables the tracing of consistent value evolutionary trajectories and the comparison of distinct cultural contexts between China and the United States~\cite{cao2024united}. To support this, we reconstruct multi-wave datasets from waves 5, 6, and 7 of the World Values Survey (WVS), stratifying populations by the aforementioned four sociodemographic dimensions, which results in 28 Chinese groups and 23 U.S. groups.

Based on the constructed dataset, we conduct a thorough analysis to model and forecast group-level value evolution in the two countries—critical for anticipating future societal shifts and understanding diverse groups’ long-term trajectories. Specifically, we propose a comprehensive simulation framework with two key steps.

First, we establish a foundational \textit{\textbf{Value Trajectory Prediction (VTP)}} model by fine-tuning LLMs to learn dynamic value transitions across survey waves. By analyzing a group’s historical responses to predict subsequent values, this approach effectively captures intrinsic evolutionary trends and subtle group-specific patterns.

Second, recognizing that external shocks drive value change~\cite{collis2022global,giuliano2025aggregate,durkheim2023rules}, we further propose an enhanced \textit{\textbf{Event-Aware Prediction (EAP)}} model. We retrieve the most semantically relevant events from a large-scale event pool by projecting event representations onto value vectors. LLMs then reason over these events to estimate their impact, enabling interpretable modeling of how historical major events shape group-level value dynamics. Our EAP model demonstrates the ability to accurately capture differences in value stability across social groups, providing meaningful insights for future social simulation and forecasting. In summary, the main contributions of this work are as follows.

\begin{itemize}
    \item We make the first attempt to integrate historical value trajectories into LLM-based human response modeling, enabling LLMs to capture the dynamic evolution of group-level values, and we release a multi-wave, group-stratified dataset for training and evaluation.

    \item We propose an event-aware simulation method that uses social values as a bridge to predict future group-level value trends. By aligning social events with value representations, our model enables interpretable analysis of event-driven value shifts.

    \item Extensive experimental results demonstrate the robustness and generalizability of our framework on both seen and unseen questions. Furthermore, the analysis reveals that U.S. groups exhibit significantly higher volatility compared to Chinese groups, thus underscoring distinct sensitivity patterns across demographic groups.
\end{itemize}

\section{Related Work}
The development of LLMs has introduced new methodologies for exploring sociological issues. Recent studies demonstrate that LLMs can serve as effective human proxies, capable of reflecting social patterns and replicating human-like survey behaviors across various contexts~\cite{argyle2023out,sun2024random,cao2025specializing}. By simulating large populations at a lower cost than traditional manual surveys, researchers use LLMs to model multidimensional sociological phenomena, including moral decision-making~\cite{ramezani2023knowledge,haemmerl2023speaking}, the preservation of cultural identities~\cite{cao2023assessing,wang2024cdeval,alkhamissi2024investigating}, and the distribution of human values across different social strata~\cite{santurkar2023whose,yao2024value,xu2025empowering}.

To improve the accuracy of human behavior simulation, current frameworks primarily rely on persona-based prompting. Specifically, early methods~\cite{cao2025specializing,liu2025towards} integrated detailed demographic labels and social identities to elicit responses consistent with pre-defined personas. However, these approaches rely on limited traits such as nationality or basic demographic categories (age and gender), and they often treat social identity as a fixed state. This static modeling fails to capture the psychological depth of real individuals, leading to limited simulation authenticity~\cite{deshpande2023toxicity}.

To address these limitations, researchers have developed advanced prompting methods that go beyond basic demographic cues to provide richer context. For instance, Alkhamissi~\cite{alkhamissi2024investigating} et al. and Hwang~\cite{hwang2023aligning} et al. proposed integrating multidimensional attributes such as social status, education, and occupation to improve grounding. Furthermore, Liu et al.~\cite{liu2025beyond} introduced an MBTI-based reasoning framework, which structures response generation into sequential cognitive steps aligned with personality traits.

While these advancements improve the modeling of group's internal state at a specific moment, they overlook the dynamic evolution of human values, failing to account for temporal changes and the impact of external events on group-level values. In contrast, we propose modeling the historical value trajectories of social groups. By incorporating these trajectories and examining the influence of major social events, our approach enables more accurate learning and simulation of dynamic value evolution.

\begin{figure*}[t]
    \centering
    \includegraphics[width=\textwidth]{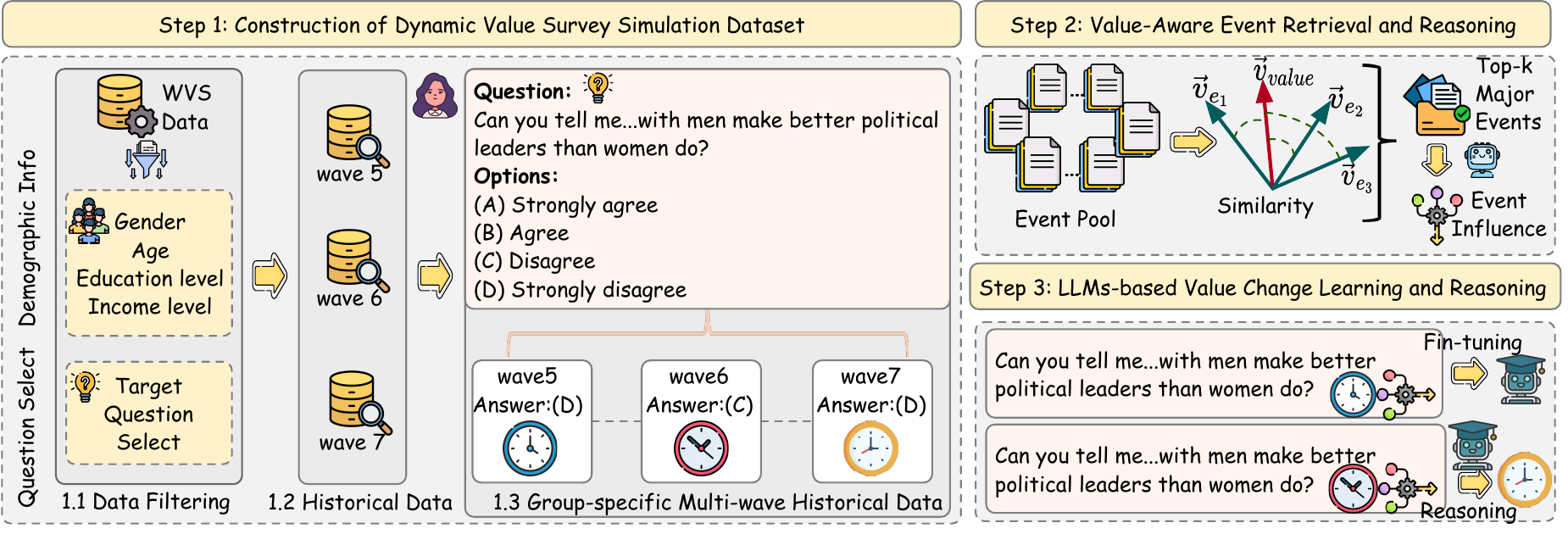}
    \caption{The overview of our proposed framework. }
    \label{fig:method}
\end{figure*}

\section{Dynamic Value Survey Simulation Dataset} \label{sec:Value Survey Simulation Dataset}
\subsection{Data Source}

We construct our dynamic simulation dataset from waves 5\footnote{\url{https://www.worldvaluessurvey.org/WVSDocumentationWV5.jsp}}, 6\footnote{\url{https://www.worldvaluessurvey.org/WVSDocumentationWV6.jsp}}, and 7\footnote{\url{https://www.worldvaluessurvey.org/WVSDocumentationWV7.jsp}} of the WVS. Our study focuses on China (2007, 2012, 2018) and the United States (2006, 2011, 2017), selected as key representatives of Eastern and Western cultures, respectively. As a leading cross-national survey, the WVS utilizes standardized questionnaires to systematically assess core human values, such as democratization, religious beliefs, and social capital. Its rigorous design ensures both cross-cultural and longitudinal comparability, providing a robust empirical foundation for analyzing value evolution over time.

To ensure data quality, we apply a filtering process to the raw responses. We exclude non-substantive entries labeled as ``No answer'', ``Don’t know'' or ``Refused'' as well as any records with missing values~\cite{cao2025specializing}. This step ensures that the retained samples reflect genuine and clearly defined value orientations, thereby enhancing the validity and reliability of our group-level value modeling.

\begin{table}[t]
\centering
\caption{The ten thematic categories of social values selected from the WVS.}
\small 
\begin{tabular}{cl}
\toprule
\textbf{ID} & \textbf{Value Thematic Category} \\
    \midrule
    1 & Social Values, Attitudes and Stereotypes \\
    2 & Happiness and Well-being \\
    3 & Social Capital, Trust and Organizational Membership \\
    4 & Economic Values \\
    5 & Security \\
    6 & Science and Technology \\
    7 & Religious Values \\
    8 & Ethical Values and Norms \\
    9 & Political Interest and Political Participation \\
    10 & Political Culture and Political Regimes \\
    \bottomrule
\end{tabular}

\label{tab:value_themes}
\end{table}

\subsection{Dynamic Data Selection}
To rigorously evaluate the model’s ability to simulate dynamic value evolution, it is essential to construct a ground-truth benchmark that faithfully reflects real historical trajectories. To this end, we build a dynamic simulation dataset that captures both stable, enduring values and shifts in public opinion on emerging societal issues across multiple survey waves. This dataset ensures semantic consistency across waves while preserving demographic representativeness, thereby enabling a reliable and fine-grained assessment of longitudinal value modeling and cross-temporal generalization.

\textbf{\textit{Dynamic Survey Question Selection.}} To systematically evaluate the model's performance, we categorize the survey items into two distinct sets: Seen Questions and Unseen Questions. This design is intended to benchmark both the model's ability to capture longitudinal value trajectories and its capacity to infer attitudes on emerging social issues.

\textbf{Seen questions} refer to the questions that are common to waves 5, 6, and 7 of the WVS. Their semantic stability ensures cross-temporal comparability, allowing us to reliably characterize group-level value evolution. Based on this criterion, we identify 98 common questions for China and 111 for the United States. These items form the foundation for evaluating the model's consistency in capturing long-term societal trends.

\textbf{Unseen questions} consist of items appearing in waves 6 and 7 but absent from wave 5. These serve as a test to evaluate whether the model can produce reasonable inferences when confronted with newly emerging social topics. This set includes 31 questions for China and 42 for the United States, providing a measure of the model's inferential robustness across shifting social discourses.

In total, the selected questions span 10 predefined WVS thematic categories, ranging from political interest to religious values, thereby covering a broad spectrum of social, cultural, and ideological dimensions. This thematic diversity ensures a comprehensive and balanced evaluation across a wide ideological landscape rather than a narrow or domain-specific focus. The detailed categorization of these value themes is presented in Table~\ref{tab:value_themes}.

\begin{table}[t]
    \centering
    \caption{The demographic dimensions used when prompting the model to simulate responses from specific survey groups.}
        \renewcommand{\arraystretch}{1.2} 
        \begin{tabular}{ll}
        \toprule
        \textbf{Dimension} & \textbf{Possible Values} \\
        \midrule
        \textbf{Sex}       & Male, Female \\
        \textbf{Age}       & \makecell[l]{Young (18-44), Middle Age (45-59), \\ Young-Old (60-74)} \\
        \textbf{Highest Education Level} & Higher, Middle, Lower \\
        \textbf{Income Level}  & High, Medium, Low \\
        \bottomrule
        \end{tabular}
        \renewcommand{\arraystretch}{1.0} 

    \label{tab:demographic1}
\end{table}

\begin{table}[t]
\centering
\caption{
    \label{tb:dataset_statistics} \textbf{WVS dataset statistics} for China and the United States across Participant-group ($P$), single-choice Question ($Q_1$) and scale Question ($Q_2$) splits.
}
\resizebox{\columnwidth}{!}{
    \begin{tabular}{l|cccc|cccc} 
        \toprule
        \multirow{2}{*}{\textbf{Split}} & \multicolumn{4}{c|}{\textbf{China}} & \multicolumn{4}{c}{\textbf{United States}} \\ 
        \cmidrule(lr){2-5} \cmidrule(lr){6-9} 
        &P & $Q_1$ & $Q_2$ & \textbf{Entries} & P & $Q_1$ & $Q_2$ & \textbf{Entries} \\ \midrule \midrule
        \textbf{Train} & 28 & 68 & 30 & $\bm{2744}$ & 23 & 80 & 31 & $\bm{2553}$ \\ \midrule
        \textbf{Test1} & 28 & 68 & 30 & $\bm{2744}$ & 23 & 80 & 31 & $\bm{2553}$ \\ \midrule
        \textbf{Test2} & 28 & 23 & 8 & $\bm{868}$  & 23 & 34 & 8 & $\bm{966}$  \\ 
        \bottomrule
    \end{tabular}
}
\end{table}

\textbf{\textit{Dynamic Survey Participant-Group Selection.}} We stratify respondents by standard dimensions in the WVS including sex, age, highest education level, and income level, for analyzing group-level value differences~\cite{bialowolski2025analysis, batruch2025advancing, elbaek2023subjective}. Gender, education level, and income level are categorized following the standard classifications provided by the WVS. Due to the wide dispersion of ages, we segment age according to the World Health Organization (WHO) 2025 guidelines~\cite{wong2025policy,who2020world} into Young, Middle-aged and Young-old groups, as detailed in Table~\ref{tab:demographic1}.

While these attributes theoretically yield 54 demographic groups, we prune underrepresented groups to guarantee statistical robustness, resulting in 28 groups for China and 23 for the United States. These groups span four key demographic dimensions, providing representative coverage of social group structures across countries. By focusing on these distinct groups, we provide a rigorous benchmark for evaluating whether the model can discern fine-grained ideological shifts across different segments of society. The group selection standard and group counts are provided in Appendix \ref{tb:demographic_stats}.

After identifying demographic groups and survey questions, we paired each group with its corresponding questions. For each group, the representative answer is determined by majority vote, selecting the option most frequently chosen by cohort members. This ensures that the responses accurately reflect each group’s general inclination, yielding a structured dataset for each country. In each survey wave, the China dataset contains 2,744 data points for seen questions and 868 for unseen questions, while the U.S. dataset includes 2,553 and 966 points, respectively. Dataset statistics are summarized in Table \ref{tb:dataset_statistics}. Test1 denotes questions that are identical to those in the training set, while Test2 denotes questions that are entirely different from those in the training set.

\subsection{Dynamic Value Prompt Formulation} \label{sec:Value-Driven Event Alignment}

To model diachronic value shifts, we design a structured dynamic longitudinal prompt that grounds LLMs in explicit socio-temporal contexts. As shown in Table~\ref{tb:questionnaire_example}, each prompt consists of four modular components.

\begin{itemize}[leftmargin=*,noitemsep,topsep=2pt]
\item \textbf{Instruction (Persona \& Temporal Grounding)}.
We encode a multi-dimensional demographic vector $\mathcal{X}_{\text{pers}}$ (country, age, gender, education and income) together with the target year $T_{\text{target}}$, anchoring the model to a specific group identity and time point.

\item \textbf{Input \& Options (Survey Specification)}.
The task content is defined by the survey question $Q$ and a constrained option set $\text{Opt}=\{o_1,\dots,o_K\}$, ensuring controlled and comparable outputs.

\item \textbf{Format (Historical Contextualization)}.
We include the historical group-level response $\mathcal{Y}_{T_{\text{prev}}}$ from the previous survey wave $T_{\text{prev}}$ as a longitudinal prior, guiding the prediction of the target response $\mathcal{Y}_{T_{\text{target}}}$.

\item \textbf{Output}.
A label-only output requirement is imposed to reduce generative noise and enable standardized evaluation.
\end{itemize}

Formally, the longitudinal prompt $P$ at time $T_{\text{target}}$ is defined as:
\begin{equation}
P_{base} = \text{Inst}(\mathcal{X}_{\text{pers}}, T_{\text{target}}) 
\oplus \text{Task}(Q, \text{Opt}) 
\oplus \text{Hist}(\mathcal{Y}_{T_{\text{prev}}}),
\end{equation}
where $\text{Inst}(\cdot)$ injects persona and temporal information, $\text{Task}(\cdot)$ specifies the survey question and options, and $\text{Hist}(\cdot)$ provides the historical prior.

The model is trained to maximize the likelihood of the correct label $y$ given $P_{base}$. We fine-tuning LLMs on the constructed dataset using the proposed longitudinal prompt formulation, thereby building a base \textbf{Value Trajectory Prediction} \textbf{(VTP)} model that learns and characterizes the dynamic evolution of group-level values across survey waves.

\begin{table}[t]
\centering
\caption{\textbf{Example entry} from our formatted WVS dataset for the country of United States.}
\label{tb:questionnaire_example}
\resizebox{0.45\textwidth}{!}{%
\small
\begin{tabular}{@{}p{0.11\textwidth} p{0.3\textwidth}@{}}
\toprule
\textbf{Instruction} & You are currently in United States in \textbf{2017}. You are a \textbf{Middle Age} \textbf{female} aged 45 to 59, have \textbf{higher} education, and your current income level is \textbf{medium}. How would you answer the following question? No thinking, no explanation. Output the answer directly. \\
\midrule
\textbf{Input} & Please indicate how important family is in your life? Please select one option from the following. \\ 
\midrule
\textbf{Options} & \text{(A) Very interested} \\
                 & \text{(B) Somewhat interested} \\
                 & \text{(C) Not very interested} \\
                 & \text{(D) Not at all interested} \\
\midrule
\textbf{Format} & Historical Answer: \newline The answer in 2011 is A. \newline
Predict the answer in 2017: \\ 
\midrule

\textbf{Output}& \text{(A)} \quad   \\           \bottomrule
\end{tabular}}

\end{table}

\section{Event-Aware Prediction Model}
To more accurately model the dynamic evolution of values and uncover the driving forces behind temporal changes, we further propose an enhanced \textbf{Event-Aware Prediction (EAP)} model built upon the VTP model. As shown in Figure~\ref{fig:method}, this framework explicitly incorporates major social events as exogenous signals for value shifts, enabling the model to capture not only the outcomes of value change but also the underlying causes. Specifically, the framework consists of three key components: the construction of an event pool, value-driven event matching, and value prediction.

\subsection{Event Pool Construction}
We construct a domain-specific, cross-temporal event pool for China and the United States, focusing on major events empirically shown to influence public opinion and social values. The construction process relies on authoritative sources, applies salience-based filtering, and aligns events with relevant values to ensure temporal consistency and sociological validity. Event information is aggregated from multiple reliable channels to guarantee societal relevance and data quality. Ultimately, this process results in two comprehensive event pools, one for China and the other for the United States.
\begin{itemize}

    \item \textbf{United States event pool.} We collect 595 significant events (2007--2010, 2012--2016\footnote{To avoid temporal ambiguity, events in WVS survey years (2006, 2011, 2016) are omit as exact collection dates are unavailable. This rule applies to China events.}) from the \textit{Pew Research Center}. Key themes, including gender, education, and the environment, are selected for their strong alignment with WVS value dimensions, ensuring sociological validity.
    
    \item \textbf{China event pool.} Because the Pew Research Center does not provide event data for China, we collect 1,107 significant events (2008--2011, 2013--2017) from \textit{Wikipedia}, \textit{China News Service}, and \textit{People's Daily}. These high-authority sources ensure comprehensive coverage of milestones resonant within the Chinese public sphere.
    
\end{itemize}

\begin{table*}[!]
\centering
\caption{Performance comparison across models and methods on China and United States datasets. The best-performing results for each model are indicated in bold. EM denotes the score for single-choice questions, PS for scale-based questions, and Overall for the overall score.}
\small
\setlength{\tabcolsep}{5pt}
\renewcommand{\arraystretch}{1.2}

\begin{tabular}{c|c|ccc|ccc|ccc|ccc}
\hline
\multirow{3}{*}{\textbf{Model}} & 
\multirow{3}{*}{\textbf{Methods}} & \multicolumn{6}{c|}{\textbf{China}} & \multicolumn{6}{c}{\textbf{United States}} \\
\cline{3-14}
& & \multicolumn{3}{c|}{Seen Question} & \multicolumn{3}{c|}{Unseen Question} & \multicolumn{3}{c|}{Seen Question} & \multicolumn{3}{c}{Unseen Question} \\
\cline{3-14}
& & $EM$ & $PS$ & \textit{Overall}  
  & $EM$ & $PS$ & \textit{Overall} 
  & $EM$ & $PS$ & \textit{Overall} 
  & $EM$ & $PS$ & \textit{Overall}  \\ 
\hline

\multirow{4}{*}{\textbf{Qwen3-8B}} 
& \multicolumn{1}{c|}{Vanilla} & 55.78 & 40.74 & 51.17 & 49.53 & 42.59 & 47.74 & 52.77 & 55.60 & 53.56 & 53.71 & 62.72 & 55.42 \\
& \multicolumn{1}{c|}{VTP} & 68.28 & 61.26 &66.13& 75.31 & 77.14 & 75.78  & 62.50 & 75.01 & 65.99  & 63.17 & 76.74 & 65.76 \\
& \multicolumn{1}{c|}{EAP} & {\textbf{80.51}} & \textbf{{80.64}} & \textbf{{80.55}}& \textbf{79.35} & \textbf{80.09} & \textbf{79.54}   & \textbf{{71.63}} & \textbf{{78.51}} & \textbf{{73.55}} & \textbf{{72.25}} & \textbf{{80.33}} & \textbf{73.79}\\ 
\cline{1-14}

\multirow{4}{*}{\textbf{Qwen3-14B}} 
& \multicolumn{1}{c|}{Vanilla} & 64.81 & 48.43 & 59.80 & 70.96 & 63.66 & 69.08 & 48.48 & 62.27 & 52.33 & 57.67 & 70.87 & 60.19 \\
& \multicolumn{1}{c|}{VTP}& 75.63 & 61.07 & 71.17& 82.76 & \textbf{83.75} & 83.02  & 66.74 & 69.93 & 67.63 & 67.65 & \textbf{{78.37}} & 69.69 \\
& \multicolumn{1}{c|}{EAP} & \textbf{{82.72}} & \textbf{{79.71}} & \textbf{{81.80}}& \textbf{84.16} & 81.16 & \textbf{83.39}  & \textbf{{77.12}} & \textbf{{79.33}} & \textbf{{77.74}} & \textbf{{70.72}} & 76.96  & \textbf{{71.90}} \\ 
\cline{1-14}

\multirow{4}{*}{\textbf{GLM4-9B}} 
& \multicolumn{1}{c|}{Vanilla} & 53.68 & 51.60 & 53.04  & 52.95 & 71.52 & 57.74  & 51.74 & 60.28 & 54.12 & 47.31 & 63.15 & 50.33 \\
& \multicolumn{1}{c|}{VTP} & 60.71 & 70.64 & 63.75& 84.94 & 82.86 & 84.40   & 74.62 & 79.80 & 76.07  & 71.87 & 74.89 & 72.44  \\
& \multicolumn{1}{c|}{EAP} & \textbf{{77.10}} & \textbf{{83.36}} & \textbf{{79.02}}& \textbf{87.42} & \textbf{85.18} & \textbf{86.84}  & \textbf{{79.18}} & \textbf{{82.97}} & \textbf{{80.24}}  & \textbf{{76.21}} & \textbf{{81.63}}  & \textbf{{77.25}} \\ 
\hline

\multirow{4}{*}{\textbf{Llama3.1-8B}} 
& \multicolumn{1}{c|}{Vanilla} & 48.16 & 42.62 & 46.47 & 49.07 & 36.34 & 45.78 & 51.68 & 55.54 & 52.76 & 46.93 & 42.93 & 46.17 \\
& \multicolumn{1}{c|}{VTP} & 53.94 & 51.17 &53.09& 75.31 & 80.18 & 76.57  & \textbf{77.77} & 80.31 & 78.48 & 59.08 & \textbf{79.02}  & 62.88\\ 
& \multicolumn{1}{c|}{EAP} & \textbf{{78.68}} & \textbf{{74.36}} & \textbf{{77.35}}& \textbf{79.35} & \textbf{80.89} & \textbf{79.75}  & 77.45 & \textbf{{81.77}} & \textbf{{78.65}}  & \textbf{{70.72}} & 74.13  & \textbf{{71.37}} \\ 
\cline{1-14}

\multirow{4}{*}{\textbf{Mistral3-7B}} 
& \multicolumn{1}{c|}{Vanilla} & 38.60 & 55.93 & 43.91 & 64.44 & 58.48 & 62.90 & 49.13 & 61.94 & 52.71 & 46.04 & 60.87 & 48.86 \\
& \multicolumn{1}{c|}{VTP} & 59.24 & {62.29} & 60.17& \textbf{70.96} & 73.84 & 71.71  & 67.50 & \textbf{71.87} & 68.72 & 59.08 & \textbf{70.76}  & 61.30 \\
& \multicolumn{1}{c|}{EAP} & \textbf{{74.11}} &\textbf{67.33} & \textbf{{72.03}}& 70.50 & \textbf{76.16} & \textbf{71.96}  & \textbf{{73.21}} & 70.13 & \textbf{{72.35}} & \textbf{{65.22}} & 70.43 & \textbf{{66.21}} \\ 
\hline

\bottomrule
\multirow{2}{*}{\textbf{GPT-4o-mini}} 
& \multicolumn{1}{c|}{Vanilla} & 55.83 & 59.69 & 57.01& 43.48 & 73.57 & 51.24  & 49.57 & 69.06 & 55.01 & 55.12 & 68.48 & 57.66 \\
& \multicolumn{1}{c|}{EAP\_noft} & \textbf{59.14} & \textbf{{72.64}} &\textbf{{63.27}}& \textbf{70.81} & \textbf{83.04} & \textbf{73.96}  & \textbf{{69.13}} & \textbf{72.26} & \textbf{{70.00}} & \textbf{70.97} & \textbf{69.67} & \textbf{70.72} \\
\cline{1-14}
\multirow{2}{*}{\textbf{Qwen-max}}
& \multicolumn{1}{c|}{Vanilla} & 64.50 & 64.57 & 64.52& 67.24 & 66.52 & 67.05  & 62.66 & 71.08& 65.01 & 66.75 & 69.89 & 67.35 \\
& \multicolumn{1}{c|}{EAP\_noft} & \textbf{70.64} & \textbf{{78.21}} & \textbf{{72.96}}& \textbf{84.32} & \textbf{82.14} & \textbf{83.76}  & \textbf{{75.92}} & \textbf{{83.73}} & \textbf{{78.10}} & \textbf{76.34} & \textbf{82.61} & \textbf{77.54} \\
\cline{1-14}
\hline
\end{tabular}

\label{tab:main_results}
\end{table*}

\subsection{Value-Driven Major Event Matching}
Public attitudes reflect core values rather than literal phrasing. For instance, ``men make better political leaders than women do'' goes beyond superficial keywords like ``men'' or ``women''. Since direct matching fails to capture these deeper sociological drivers, we propose \textbf{Value-Driven Major Event Matching}, which employs value vectors as a semantic bridge, prioritizing alignment with the underlying values that fundamentally shape public opinion.

We begin with a value annotation foundation built on the WVS. Each survey question $q \in \mathcal{Q}$ is explicitly tagged with its corresponding value dimension $v \in \mathcal{V}$ (``Religious Values'', ``Security'') based on the official WVS codebook. This mapping can be defined as $v = \mathcal{M}(q)$, where $\mathcal{M}$ is the assignment function. For example, questions measuring agreement with ``men make better political leaders than women do'' are mapped to the Stereotypes dimension. This creates a structured question-value base where every survey item is represented not by its wording, but by its underlying value label $v$.

With the value dimensions for each survey question anchored, we proceed to encoding using domain-adaptive pre-trained language models. To ensure cross-lingual robustness, we adopt a bilingual embedding strategy. For a value dimension $v$, its embedding vector $\mathbf{v}$ is generated by:
\begin{equation}
\mathbf{v} = E(v), \quad E \in \{E_{zh}, E_{en}\} ,
\end{equation}
where $E_{zh}$ and $E_{en}$ denote bge-large-zh-v1.5 and bge-large-en-v1.5, respectively. Similarly, we perform event encoding to map each event $e$ in the event set $\mathcal{E}$ into the same vector space, yielding event vector $\mathbf{e} = E(e)$. Finally, we execute semantic similarity matching by calculating the cosine similarity between each value vector $\mathbf{v}$ and event vector $\mathbf{e}$:
\begin{equation}
\text{sim}(v, e) = \frac{\mathbf{v} \cdot \mathbf{e}}{\|\mathbf{v}\| \|\mathbf{e}\|}
\end{equation}
We then identify the top-k most aligned events $\mathcal{E}^*$ for each value dimension by sorting the similarity scores:
\begin{equation}
\mathcal{E}^* = \text{arg Top-k}_{e \in \mathcal{E}} \left( \text{sim}(v, e) \right)
\end{equation}
This pipeline transforms abstract value labels into computable semantic representations, enabling a robust, text-agnostic alignment.

Based on the retrieved top-$k$ events $\mathcal{E}^*$, our framework analyzes their potential influence $\mathcal{I}$ on specific demographic groups $G$ regarding the target questions $q$ within a timeframe $T$:
\begin{equation}
\mathcal{I} = f(\mathcal{E}^*, G, q, T) ,
\end{equation}
where $f(\cdot)$ represents the inference module. These inferred impact levels are subsequently synthesized and integrated into the final structural prompt $P = \text{Concat}(P_{base}, \mathcal{I})$, providing enriched contextual grounding to enhance the accuracy and relevance of downstream reasoning tasks. The detailed prompts used to guide event impact analysis are provided in Appendix \ref{event_impact}.

\subsection{Value Prediction}
Once event impact signals are integrated into the prompt, we train LLMs on the simulated dataset from §\ref{sec:Value-Driven Event Alignment} to capture value changes across consecutive survey waves. During inference, the model leverages wave 6 inputs, longitudinal value trajectories, and event signals, allowing it to predict wave 7 responses with improved accuracy and a faithful representation of group-level value dynamics. Examples of the final reasoning prompts in Chinese and English are provided in the Appendix \ref{prompt}.

\section{Experiment}
\subsection{Evaluation Models}
To facilitate a comparative analysis of values between China and the United States, our selection specifically balances models with Chinese and English linguistic backgrounds. We fine-tuned five open-source LLMs from four representative families: Qwen3-8B~\cite{yang2025qwen3}, Qwen3-14B~\cite{yang2025qwen3}, Llama3.1-8B-Instruct~\cite{grattafiori2024llama}, GLM4-9B-0414~\cite{glm2024chatglm} and Mistral3-7B-Instruct-V0.3~\cite{jiang2023mistral7b}. By including models in Chinese (Qwen and GLM) and English (Llama and Mistral) corpora, we aim to ensure the framework's robustness in capturing cross-cultural and cross-linguistic value evolutions. The inclusion of Qwen3 variants allows us to observe potential scaling effects. Additionally, GPT-4o-mini~\cite{achiam2023gpt} and Qwen-Max~\cite{yang2025qwen3} are included as representative English and Chinese-centric closed-source model. All open-source models are fine-tuned following the protocol in §\ref{sec:Value-Driven Event Alignment}. Further details on models and inference setup are provided in Appendix \ref{setup}.

\subsection{Baselines}
To better evaluate the effectiveness of our approach in modeling value dynamics, we compare three paradigms: \textbf{Vanilla}, Value Trajectory Prediction (\textbf{VTP}) and Event-Aware Prediction (\textbf{EAP}). In Vanilla, the model directly predicts responses based on group features alone. Both VTP and EAP adopt Low-Rank Adaptation (LoRA)~\cite{hu2022lora} for fine-tuning on waves 5–6 group data. While VTP leverages only historical value trajectories, EAP further incorporates major event impact signals to enhance prediction. During inference, wave 6 group values are used to predict wave 7 outcomes. EAP\_noft denotes inference on closed-source models using historical value trajectories and event signals, without fine-tuning.

\subsection{Metrics}
To better evaluate the alignment between model simulations and real-world data, we adopt three metrics. For single-choice questions, we use Exact Match (\textbf{EM}). While for scaled-response questions, we employ the Proximity Score (\textbf{PS}). The PS reflects the model’s sensitivity to fine-grained value nuances, assigning partial credit to 1–10 scale questions within a distance of 5 to capture sentiment intensity~\cite{saris2014design,russell1980circumplex}. Finally, an Overall proportional score (\textbf{Overall}) is computed by weighting these metrics according to the proportion of each question type, providing a comprehensive measure of the model’s holistic performance.


\subsection{Main Results}\label{sec:main results}
We evaluate three types of methods, including  the Vanilla, and our proposed VTP and EAP,  on both seen and unseen questions to assess their robustness and generalization capability. Table~\ref{tab:main_results} provides the results across social groups in both China and the United States by adopting 7 LLMs as evaluation models.  Below are main results.

\textit{\textbf{Vanilla vs.\ VTP and EAP.}}
Compared with the vanilla baseline, both VTP and EAP lead to consistent performance improvements. Moreover, compared with VTP, EAP generally achieves better performance, demonstrating that the integration of event-aware signals further helps the model adjust to contextual shifts, improving its robustness capability. 

For the Chinese groups, EAP achieves the best performance on seen questions fine-tuned Qwen3-14B, improving by 22.00\% over the vanilla baseline. For the U.S. groups, EAP achieves the best performance on seen questions fine-tuned GLM4-9B, with a 26.12\% improvement over the vanilla baseline. This demonstrates the effectiveness of our EAP, which can capture both historical value trajectories and event influences, enabling the model to learn group-specific patterns more accurately and produce coherent value predictions across different demographic segments.

\textit{\textbf{Open-source Models vs.\ Closed-source Models.}}
Moreover, compared with closed-source models, fine-tuning smaller open-source LLMs can match or even surpass the performance of closed-source models. These substantial improvements further indicate that our method enables LLMs to more effectively capture group-specific value trajectories and the underlying drivers of their evolution, thereby better characterizing social groups and achieving more accurate social simulations.

For the Qwen3 series models, Qwen3-14B outperforms Qwen3-8B, indicating that increasing model scale enhances performance on social value prediction tasks and allows larger models to better capture complex shifts in group values. Notably, the Qwen3 series achieves significant gains even without fine-tuning, a capability absent in other model families. This likely stems from its strong reasoning ability, enabling effective use of historical value trajectories during inference.

Overall, the comprehensive analysis across all LLMs demonstrates that VTP and EAP methods consistently achieve strong performance, regardless of the underlying base model, indicating their broad applicability and stability. Moreover, they perform well for both Chinese and U.S. groups, suggesting that the proposed framework effectively captures universal as well as culture-specific value patterns. This robustness further highlights their potential for large-scale cross-cultural social simulation.

\textit{\textbf{Seen Questions \& Unseen Questions.}}
We further assess model generalization by evaluating performance on unseen questions. Across all models and settings, unseen questions pose a greater challenge than seen ones, as they more faithfully reflect whether the LLM has genuinely learned group-specific value representations, whether it can respond appropriately based on the characteristics of each social group.

As shown in Table~\ref{tab:main_results}, the results on unseen questions for both Chinese and U.S. groups indicate that EAP method achieves strong simulation performance even on previously unseen issues. These findings suggest that our EAP method enables the model to effectively capture group-specific value tendencies, allowing it to generate coherent simulations for unseen questions based on learned group characteristics.

Interestingly, for Chinese groups, performance on unseen questions occasionally exceeds that on seen ones. This may be because certain value changes are relatively stable across survey waves, allowing the EAP model to generalize learned tendencies to new but semantically similar questions. This phenomenon highlights the EAP's ability to leverage stable group-level patterns to enhance generalization.

\begin{table}[t]
    \centering
    \caption{Ablation results on overall metrics across five different LLMs. We compare the full method (Ours) against variants without event impact (w/o event) and without both history and event impact (w/o history\&event).}
    \label{tab:ablation_all}
    \resizebox{\columnwidth}{!}{ 
    \begin{tabular}{ll|cc|cc}
        \toprule
        \multirow{2}{*}{\textbf{Model}} & \multirow{2}{*}{\textbf{Methods}} & \multicolumn{2}{c|}{\textbf{China}} & \multicolumn{2}{c}{\textbf{United States}} \\
        \cmidrule(lr){3-4} \cmidrule(lr){5-6}
         & & Seen & Unseen & Seen & Unseen \\
        \midrule
        \midrule

        \multirow{3}{*}{\textbf{Qwen3-8B}} 
          & Ours & \textbf{74.43} & \textbf{76.34} & \textbf{69.42} & \textbf{69.63} \\
          & w/o event & 64.06 & 73.82 & 63.16 & 62.07 \\
          & w/o history\&event & 51.17 & 47.74 & 53.56 & 55.42 \\
        \midrule
        
        \multirow{3}{*}{\textbf{Qwen3-14B}} 
          & Ours & \textbf{72.19} & \textbf{82.35} & \textbf{66.61} & \textbf{70.66} \\
          & w/o event & 67.26 & 81.08 & 64.13 & 68.96 \\
          & w/o history\&event & 59.80 & 69.08 & 52.33 & 60.19 \\
        \midrule
        
        \multirow{3}{*}{\textbf{GLM4-9B}} 
          & Ours & \textbf{55.36} & \textbf{63.32} & \textbf{60.94} & \textbf{61.33} \\
          & w/o event & 54.54 & 61.47 & 58.01 & 50.85 \\
          & w/o history\&event & 53.04 & 57.74 & 54.12 & 50.33 \\
        \midrule

        \multirow{3}{*}{\textbf{Llama3.1-8B}} 
          & Ours & 45.23 & \textbf{50.62} & \textbf{63.09} & \textbf{60.08} \\
          & w/o event & 39.82 & 39.59 & 44.45 & 46.15 \\
          & w/o history\&event & \textbf{46.47} & 45.78 & 52.76 & 46.17 \\
        \midrule

        \multirow{3}{*}{\textbf{Mistral3-7B}} 
          & Ours & \textbf{60.51} & \textbf{68.82} & \textbf{62.88} & \textbf{58.53} \\
          & w/o event & 51.82 & 62.60 & 52.87 & 50.06 \\
          & w/o history\&event & 43.91 & 62.90 & 52.71 & 48.86 \\
        
        \bottomrule
    \end{tabular}
    }
\end{table}

\subsection{Ablation Study}
We conduct ablation studies across five different models to analyze the impact of each component in the framework. Since the effectiveness of fine-tuning has been demonstrated in §\ref{sec:main results}, this section focuses solely on analyzing the relative importance of different components during the inference stage. Since events without specified historical value trajectories are not meaningful for modeling value evolution, our ablation study focuses on the w/o event and w/o history\&event settings.

As shown in Table \ref{tab:ablation_all}, for both Chinese and U.S. groups, the performance of most models consistently degrades when the event impact module is removed, and the decline becomes more pronounced when both the event impact and historical trajectory modules are excluded. This indicates that these components provide complementary information for modeling group behavior. Overall, the results further demonstrate that our method effectively facilitates the learning and integration of group-level value dynamics during the inference stage, leading to more accurate and stable simulations across different populations.

Notably, across all models, the ablation of the event module leads to a more pronounced performance decline for U.S. groups than for Chinese groups, suggesting that U.S. populations are more susceptible to external events. Conversely, Chinese groups exhibit more stable value trajectories. This divergence underscores the importance of incorporating cultural context in modeling value dynamics.

Moreover, the performance of Llama3.1-8B model does not degrade after removing certain components. Further inspection of the model outputs reveals that some predictions are replaced by refusal responses (``I cannot answer this question''), which may be attributed to the model’s internal safety constraints or inherent resistance to specific inputs.

\begin{figure}[ht]
    \centering
    \includegraphics[width=0.8\columnwidth]{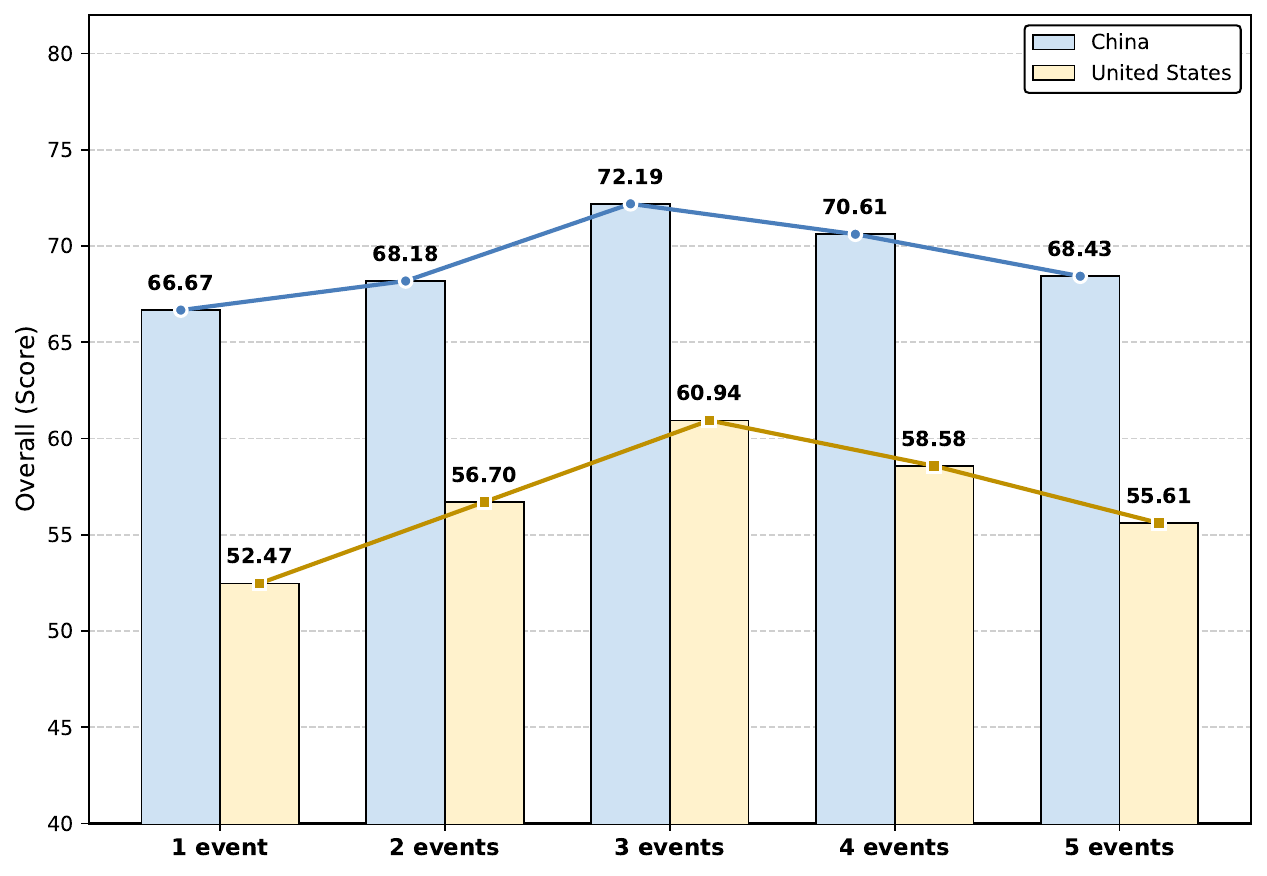}
    \caption{Experimental analysis of varying numbers of selected events for China and the United States.}
    \label{fig:us_change}
\end{figure}

\subsection{Event Impact Analysis}
As shown in Figure \ref{fig:us_change}, we examine the effect of the number of selected events on model performance using Qwen3-14B and GLM4-9B, which achieves the best results for both populations. Performance improves as the number of events increases, peaking at three, and then declines when four or five events are included.

These results indicate that a single event is insufficient to capture the group's social context, whereas selecting too many introduces noise and dilutes signals. With two events, potential conflicts may introduce ambiguity and impair judgment. In contrast, selecting three events strikes a balance between information sufficiency and coherence, leading to the optimal simulation performance.

\begin{table}[t]
\centering
\caption{\label{tb:table22} \textbf{Comparison of random event selection (Random) and direct question-event matching (Direct).}}
\resizebox{0.98\columnwidth}{!}{
\begin{tabular}{l|ccc|ccc}
\toprule
\multirow{2}{*}{Methods} 
& \multicolumn{3}{c|}{China} 
& \multicolumn{3}{c}{United States} \\
\cmidrule(lr){2-4} \cmidrule(lr){5-7}
& $EM$ & $PS$ & $Overall$ & $EM$ & $PS$ & $Overall$ \\ 
\midrule \midrule
\textbf{Ours}   
& \textbf{72.64} & \textbf{71.17} & \textbf{72.19} 
& \textbf{56.74} & \textbf{71.78} & \textbf{60.94} \\
\midrule
Random & 44.49 & 61.21 & 49.61 & 36.63 & 62.61 & 43.89 \\
Direct & 53.10 & 69.67 & 58.17 & 40.16 & 62.86 & 46.50 \\
\bottomrule
\end{tabular}}
\end{table}

In addition, we compare our event selection strategy with two baselines: randomly selecting three events and directly matching questions to events without social values. As shown in Table \ref{tb:table22}, our approach consistently outperforms both, demonstrating the necessity of value-mediated event selection for social simulation.

To better understand how events influence group-level value dynamics, we further conduct case studies to examine the impacts of specific events on value changes across different national populations. For instance, in the U.S., the event ``World Economic Forum'' exerts the strongest influence on the value shift related to economic. In China, the event ``The 19th National Congress of China'' prominently affects perceptions of social development. The corresponding case studies can be found in the Appendix \ref{case}.

\subsection{Value Change Analysis Across Groups}
\begin{figure}[ht]
    \centering
    \includegraphics[width=1\columnwidth]{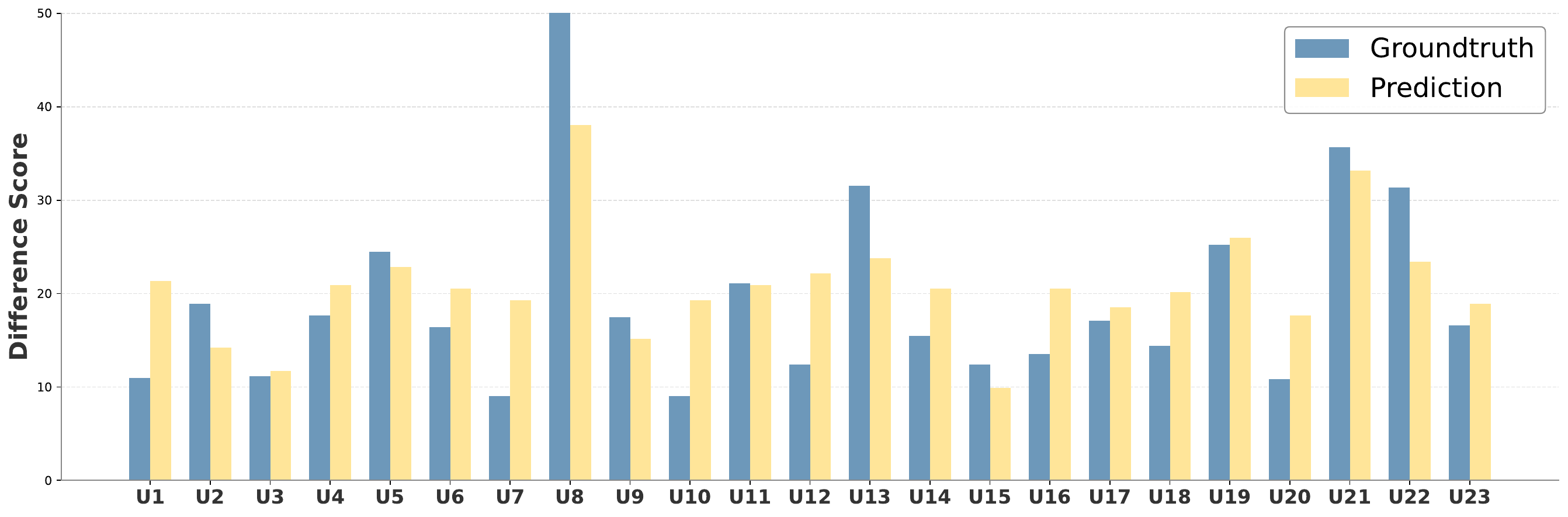}
    \caption{Changes in the 23 U.S. social groups from wave 6 to wave 7 by comparing predicted and observed values.}
    \label{fig:us_change1}
\end{figure}

\begin{figure}[ht]
    \centering
    \includegraphics[width=1\columnwidth]{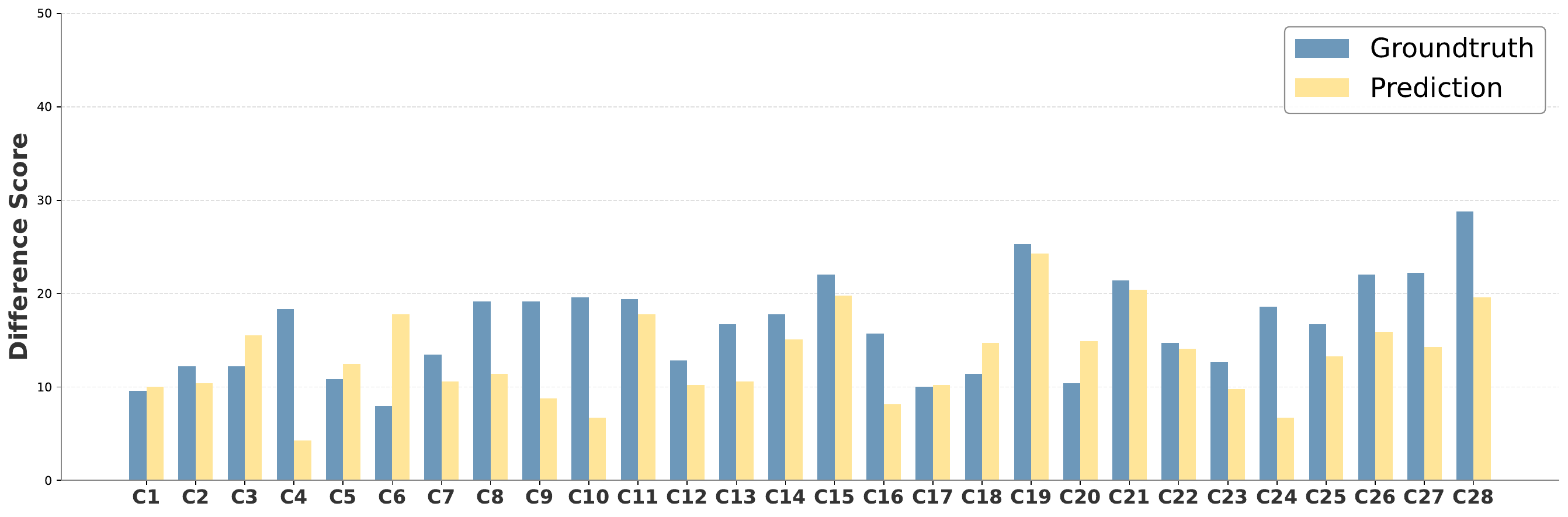}
    \caption{Changes in the 28 China social groups from wave 6 to wave 7 by comparing predicted and observed values.}
    \label{fig:us_change2}
\end{figure}

We further conduct analysis of the simulation results for both Chinese and U.S. populations. As shown in Figures \ref{fig:us_change1} and \ref{fig:us_change2}, we observe that U.S. groups generally exhibit a higher degree of value volatility compared to their Chinese counterparts. Validating these simulations against ground truth survey data further confirms our model's ability to accurately capture value shifts across diverse demographics. These findings also highlight significant cross-national heterogeneity in value stability.

A fine-grained examination identifies specific subpopulations driving these trends. In the U.S., the group exhibiting the most substantial value shift is \textit{Young, Medium-educated, High-income Female} (U8). In contrast, the most substantial value shift in China occurs in \textit{Young, Middle-educated, Medium-income Male} (C28). We also observe that in both China and the United States, younger groups tend to exhibit higher sensitivity to value changes. These insights can guide more targeted attention to sensitive demographic groups in future research.

\section{Conclusion}
In this paper, we address the limitations of static social simulation by proposing a dynamic framework that integrates historical value trajectories and major societal events. Based on our constructed multi-wave  dataset from World Values Survey for China and the United States, our approach enables LLMs to model the continuous evolution of group-level beliefs. Our event-aware prediction mechanism aligns societal shocks with internal value shifts, allowing interpretable simulations of event impacts across demographic groups. Experiments on five LLMs show superior performance on both seen and unseen questions, and analysis of value sensitivity across groups validates the model’s sociological plausibility.

\section{Limitations and Ethical Considerations}
Despite its promising results, the proposed framework aligns societal events with value shifts based on semantic relevance, it primarily captures correlational relationships, and unobserved confounding factors driving social value evolution may remain unmodeled.

We adhere to strict ethical standards in developing and applying our social simulation framework. All data are publicly available. This paper is intended solely for sociological research, policy analysis, and understanding social dynamics. We caution against misuse such as disinformation or social manipulation.

\bibliographystyle{ACM-Reference-Format}
\bibliography{reference}

@article{bialowolski2025analysis,
  title={Analysis of demographic variation and childhood correlates of financial well-being across 22 countries},
  author={Bialowolski, Piotr and Makridis, Christos A and Bradshaw, Matt and Weziak-Bialowolska, Dorota and Gundersen, Craig and Le Pertel, No{\'e}mie and Gibson, Cristina and Jang, Sung Joon and Padgett, R Noah and Johnson, Byron R and others},
  journal={Nature Human Behaviour},
  pages={1--16},
  year={2025},
  publisher={Nature Publishing Group UK London}
}

@article{batruch2025advancing,
  title={Advancing the psychology of social class with large-scale replications in four countries},
  author={Batruch, Anatolia and Sommet, Nicolas and Autin, Fr{\'e}d{\'e}rique},
  journal={Nature Human Behaviour},
  pages={1--22},
  year={2025},
  publisher={Nature Publishing Group UK London}
}

@article{elbaek2023subjective,
  title={Subjective socioeconomic status and income inequality are associated with self-reported morality across 67 countries},
  author={Elb{\ae}k, Christian T and Mitkidis, Panagiotis and Aar{\o}e, Lene and Otterbring, Tobias},
  journal={Nature Communications},
  volume={14},
  number={1},
  pages={5453},
  year={2023},
  publisher={Nature Publishing Group UK London}
}

@article{collis2022global,
  title={Global survey on COVID-19 beliefs, behaviours and norms},
  author={Collis, Avinash and Garimella, Kiran and Moehring, Alex and Rahimian, M Amin and Babalola, Stella and Gobat, Nina H and Shattuck, Dominick and Stolow, Jeni and Aral, Sinan and Eckles, Dean},
  journal={Nature Human Behaviour},
  volume={6},
  number={9},
  pages={1310--1317},
  year={2022},
  publisher={Nature Publishing Group UK London}
}

@article{giuliano2025aggregate,
  title={Aggregate shocks and the formation of preferences and beliefs},
  author={Giuliano, Paola and Spilimbergo, Antonio},
  journal={Journal of Economic Literature},
  volume={63},
  number={2},
  pages={542--597},
  year={2025},
  publisher={American Economic Association 2014 Broadway, Suite 305, Nashville, TN 37203-2425}
}

@article{glm2024chatglm,
  title={Chatglm: A family of large language models from glm-130b to glm-4 all tools},
  author={GLM, Team and Zeng, Aohan and Xu, Bin and Wang, Bowen and Zhang, Chenhui and Yin, Da and Zhang, Dan and Rojas, Diego and Feng, Guanyu and Zhao, Hanlin and others},
  journal={arXiv preprint arXiv:2406.12793},
  year={2024}
}

@article{grattafiori2024llama,
  title={The llama 3 herd of models},
  author={Grattafiori, Aaron and Dubey, Abhimanyu and Jauhri, Abhinav and Pandey, Abhinav and Kadian, Abhishek and Al-Dahle, Ahmad and Letman, Aiesha and Mathur, Akhil and Schelten, Alan and Vaughan, Alex and others},
  journal={arXiv preprint arXiv:2407.21783},
  year={2024}
}

@article{yang2025qwen3,
  title={Qwen3 technical report},
  author={Yang, An and Li, Anfeng and Yang, Baosong and Zhang, Beichen and Hui, Binyuan and Zheng, Bo and Yu, Bowen and Gao, Chang and Huang, Chengen and Lv, Chenxu and others},
  journal={arXiv preprint arXiv:2505.09388},
  year={2025}
}

@article{achiam2023gpt,
  title={Gpt-4 technical report},
  author={Achiam, Josh and Adler, Steven and Agarwal, Sandhini and Ahmad, Lama and Akkaya, Ilge and Aleman, Florencia Leoni and Almeida, Diogo and Altenschmidt, Janko and Altman, Sam and Anadkat, Shyamal and others},
  journal={arXiv preprint arXiv:2303.08774},
  year={2023}
}

@article{jiang2023mistral7b,
      title={Mistral 7B}, 
      author={Albert Q. Jiang and Alexandre Sablayrolles and Arthur Mensch and Chris Bamford and Devendra Singh Chaplot and Diego de las Casas and Florian Bressand and Gianna Lengyel and Guillaume Lample and Lucile Saulnier and Lélio Renard Lavaud and Marie-Anne Lachaux and Pierre Stock and Teven Le Scao and Thibaut Lavril and Thomas Wang and Timothée Lacroix and William El Sayed},
      year={2023},
      eprint={2310.06825},
      archivePrefix={arXiv},
      primaryClass={cs.CL}
}

@article{hu2022lora,
  title={Lora: Low-rank adaptation of large language models.},
  author={Hu, Edward J and Shen, Yelong and Wallis, Phillip and Allen-Zhu, Zeyuan and Li, Yuanzhi and Wang, Shean and Wang, Lu and Chen, Weizhu and others},
  journal={ICLR},
  volume={1},
  number={2},
  pages={3},
  year={2022}
}

@book{saris2014design,
  title={Design, evaluation, and analysis of questionnaires for survey research},
  author={Saris, Willem E and Gallhofer, Irmtraud N},
  year={2014},
  publisher={John Wiley \& Sons}
}

@article{russell1980circumplex,
  title={A circumplex model of affect.},
  author={Russell, James A},
  journal={Journal of personality and social psychology},
  volume={39},
  number={6},
  pages={1161},
  year={1980},
  publisher={American Psychological Association}
}

@article{argyle2023out,
  title={Out of one, many: Using language models to simulate human samples},
  author={Argyle, Lisa P and Busby, Ethan C and Fulda, Nancy and Gubler, Joshua R and Rytting, Christopher and Wingate, David},
  journal={Political Analysis},
  volume={31},
  number={3},
  pages={337--351},
  year={2023},
  publisher={Cambridge University Press}
}

@inproceedings{santurkar2023whose,
  title={Whose opinions do language models reflect?},
  author={Santurkar, Shibani and Durmus, Esin and Ladhak, Faisal and Lee, Cinoo and Liang, Percy and Hashimoto, Tatsunori},
  booktitle={International Conference on Machine Learning},
  pages={29971--30004},
  year={2023},
  organization={PMLR}
}

@inproceedings{cao2025specializing,
  title={Specializing Large Language Models to Simulate Survey Response Distributions for Global Populations},
  author={Cao, Yong and Liu, Haijiang and Arora, Arnav and Augenstein, Isabelle and R{\"o}ttger, Paul and Hershcovich, Daniel},
  booktitle={Proceedings of the 2025 Conference of the Nations of the Americas Chapter of the Association for Computational Linguistics: Human Language Technologies (Volume 1: Long Papers)},
  pages={3141--3154},
  year={2025}
}

@inproceedings{liu2025beyond,
  title={Beyond Demographics: Enhancing Cultural Value Survey Simulation with Multi-Stage Personality-Driven Cognitive Reasoning},
  author={Liu, Haijiang and Li, Qiyuan and Gao, Chao and Cao, Yong and Xu, Xiangyu and Wu, Xun and Hershcovich, Daniel and Gu, Jinguang},
  booktitle={Proceedings of the 2025 Conference on Empirical Methods in Natural Language Processing},
  pages={18417--18439},
  year={2025}
}

@inproceedings{alkhamissi2024investigating,
  title={Investigating Cultural Alignment of Large Language Models},
  author={Alkhamissi, Badr and ElNokrashy, Muhammad and Alkhamissi, Mai and Diab, Mona},
  booktitle={Proceedings of the 62nd Annual Meeting of the Association for Computational Linguistics (Volume 1: Long Papers)},
  pages={12404--12422},
  year={2024}
}

@article{liu2025towards,
  title={Towards realistic evaluation of cultural value alignment in large language models: Diversity enhancement for survey response simulation},
  author={Liu, Haijiang and Cao, Yong and Wu, Xun and Qiu, Chen and Gu, Jinguang and Liu, Maofu and Hershcovich, Daniel},
  journal={Information Processing \& Management},
  volume={62},
  number={4},
  pages={104099},
  year={2025},
  publisher={Elsevier}
}

@article{sun2024random,
  title={Random silicon sampling: Simulating human sub-population opinion using a large language model based on group-level demographic information},
  author={Sun, Seungjong and Lee, Eungu and Nan, Dongyan and Zhao, Xiangying and Lee, Wonbyung and Jansen, Bernard J and Kim, Jang Hyun},
  journal={arXiv preprint arXiv:2402.18144},
  year={2024}
}

@inproceedings{cao2023assessing,
  title={Assessing Cross-Cultural Alignment between ChatGPT and Human Societies: An Empirical Study},
  author={Cao, Yong and Zhou, Li and Lee, Seolhwa and Cabello, Laura and Chen, Min and Hershcovich, Daniel},
  booktitle={Proceedings of the First Workshop on Cross-Cultural Considerations in NLP (C3NLP)},
  pages={53--67},
  year={2023}
}

@inproceedings{wang2024cdeval,
  title={Cdeval: A benchmark for measuring the cultural dimensions of large language models},
  author={Wang, Yuhang and Zhu, Yanxu and Kong, Chao and Wei, Shuyu and Yi, Xiaoyuan and Xie, Xing and Sang, Jitao},
  booktitle={Proceedings of the 2nd Workshop on Cross-Cultural Considerations in NLP},
  pages={1--16},
  year={2024}
}

@inproceedings{ramezani2023knowledge,
  title={Knowledge of cultural moral norms in large language models},
  author={Ramezani, Aida and Xu, Yang},
  booktitle={Proceedings of the 61st Annual Meeting of the Association for Computational Linguistics (Volume 1: Long Papers)},
  pages={428--446},
  year={2023}
}

@inproceedings{yao2024value,
  title={Value FULCRA: Mapping large language models to the multidimensional spectrum of basic human value},
  author={Yao, Jing and Yi, Xiaoyuan and Gong, Yifan and Wang, Xiting and Xie, Xing},
  booktitle={Proceedings of the 2024 Conference of the North American Chapter of the Association for Computational Linguistics: Human Language Technologies (Volume 1: Long Papers)},
  pages={8762--8785},
  year={2024}
}

@inproceedings{haemmerl2023speaking,
  title={Speaking Multiple Languages Affects the Moral Bias of Language Models},
  author={Haemmerl, Katharina and Deiseroth, Bjoern and Schramowski, Patrick and Libovick{\`y}, Jind{\v{r}}ich and Rothkopf, Constantin A and Fraser, Alexander and Kersting, Kristian},
  booktitle={The 61st Annual Meeting Of The Association For Computational Linguistics},
  year={2023}
}

@inproceedings{deshpande2023toxicity,
  title={Toxicity in chatgpt: Analyzing persona-assigned language models},
  author={Deshpande, Ameet and Murahari, Vishvak and Rajpurohit, Tanmay and Kalyan, Ashwin and Narasimhan, Karthik},
  booktitle={Findings of the Association for Computational Linguistics: EMNLP 2023},
  pages={1236--1270},
  year={2023}
}

@inproceedings{hwang2023aligning,
  title={Aligning language models to user opinions},
  author={Hwang, EunJeong and Majumder, Bodhisattwa and Tandon, Niket},
  booktitle={Findings of the Association for Computational Linguistics: EMNLP 2023},
  pages={5906--5919},
  year={2023}
}

@article{park2024diminished,
  title={Diminished diversity-of-thought in a standard large language model},
  author={Park, Peter S and Schoenegger, Philipp and Zhu, Chongyang},
  journal={Behavior Research Methods},
  volume={56},
  number={6},
  pages={5754--5770},
  year={2024},
  publisher={Springer}
}

@inproceedings{xu2024wizardlm,
  title={WizardLM: Empowering large pre-trained language models to follow complex instructions},
  author={Xu, Can and Sun, Qingfeng and Zheng, Kai and Geng, Xiubo and Zhao, Pu and Feng, Jiazhan and Tao, Chongyang and Lin, Qingwei and Jiang, Daxin},
  booktitle={The Twelfth International Conference on Learning Representations},
  year={2024}
}

@article{mesoudi2008multiple,
  title={The multiple roles of cultural transmission experiments in understanding human cultural evolution},
  author={Mesoudi, Alex and Whiten, Andrew},
  journal={Philosophical Transactions of the Royal Society B: Biological Sciences},
  volume={363},
  number={1509},
  pages={3489--3501},
  year={2008},
  publisher={The Royal Society London}
}

@book{cavalli1981cultural,
  title={Cultural transmission and evolution: A quantitative approach},
  author={Cavalli-Sforza, Luigi Luca and Feldman, Marcus W},
  number={16},
  year={1981},
  publisher={Princeton University Press}
}

@article{mesoudi2006towards,
  title={Towards a unified science of cultural evolution},
  author={Mesoudi, Alex and Whiten, Andrew and Laland, Kevin N},
  journal={Behavioral and brain sciences},
  volume={29},
  number={4},
  pages={329--347},
  year={2006},
  publisher={Cambridge University Press}
}

@article{hills2019historical,
  title={Historical analysis of national subjective wellbeing using millions of digitized books},
  author={Hills, Thomas T and Proto, Eugenio and Sgroi, Daniel and Seresinhe, Chanuki Illushka},
  journal={Nature human behaviour},
  volume={3},
  number={12},
  pages={1271--1275},
  year={2019},
  publisher={Nature Publishing Group UK London}
}

@incollection{durkheim2023rules,
  title={The rules of sociological method},
  author={Durkheim, Emile},
  booktitle={Social theory re-wired},
  pages={9--14},
  year={2023},
  publisher={Routledge}
}

@inproceedings{xu2025empowering,
  title={Empowering Economic Simulation for Massively Multiplayer Online Games through Generative Agent-Based Modeling},
  author={Xu, Bihan and Zhao, Shiwei and Wu, Runze and Huang, Zhenya and Wang, Jiawei and Hu, Zhipeng and Wang, Kai and Liu, Haoyu and Lv, Tangjie and Li, Le and others},
  booktitle={Proceedings of the 31st ACM SIGKDD Conference on Knowledge Discovery and Data Mining V. 2},
  pages={3366--3377},
  year={2025}
}

@article{cao2024united,
  title={United States--China differences in cognition and perception across 12 tasks: Replicability, robustness, and within-culture variation.},
  author={Cao, Anjie and Carstensen, Alexandra and Gao, Shan and Frank, Michael C},
  journal={Journal of Experimental Psychology: General},
  year={2024},
  publisher={American Psychological Association}
}

@article{wong2025policy,
  title={Policy implications of WHO’s global traditional medicine strategy 2025--2034},
  author={Wong, Yuk Ming Alice and Ahn, Sangyoung and Bana, Aditi and Dua, Pradeep Kumar and Eggers, Rudi and Kuruvilla, Shyama and Li, Yachan and Liu, Qin and Shen, Yunhui and Kim, Sungchol},
  journal={Bulletin of the World Health Organization},
  volume={103},
  number={11},
  pages={715},
  year={2025}
}

@article{who2020world,
  title={World health organization},
  author={WHO, CONSTITUTION OF},
  journal={Air Quality Guidelines for Europe},
  number={91},
  year={2020}
}

@article{richiardi2006common,
  title={A common protocol for agent-based social simulation},
  author={Richiardi, Matteo G and Leombruni, Roberto and Saam, Nicole J and Sonnessa, Michele},
  journal={Journal of artificial societies and social simulation},
  volume={9},
  year={2006}
}

@article{edelmann2020computational,
  title={Computational social science and sociology},
  author={Edelmann, Achim and Wolff, Tom and Montagne, Danielle and Bail, Christopher A},
  journal={Annual review of sociology},
  volume={46},
  number={1},
  pages={61--81},
  year={2020},
  publisher={Annual Reviews}
}

@article{harrison2007simulation,
  title={Simulation modeling in organizational and management research},
  author={Harrison, J Richard and Lin, Zhiang and Carroll, Glenn R and Carley, Kathleen M},
  journal={Academy of management review},
  volume={32},
  number={4},
  pages={1229--1245},
  year={2007},
  publisher={Academy of Management Briarcliff Manor, NY 10510}
}

@article{davis1993general,
  title={General social survey},
  author={Davis, James A},
  journal={NSF Award},
  volume={91},
  number={9122462},
  pages={22462},
  year={1993}
}

\appendix
\section{Demographic Statistics} \label{tb:demographic_stats}
\begin{figure}[ht]
    \centering
    \includegraphics[width=0.8\columnwidth]{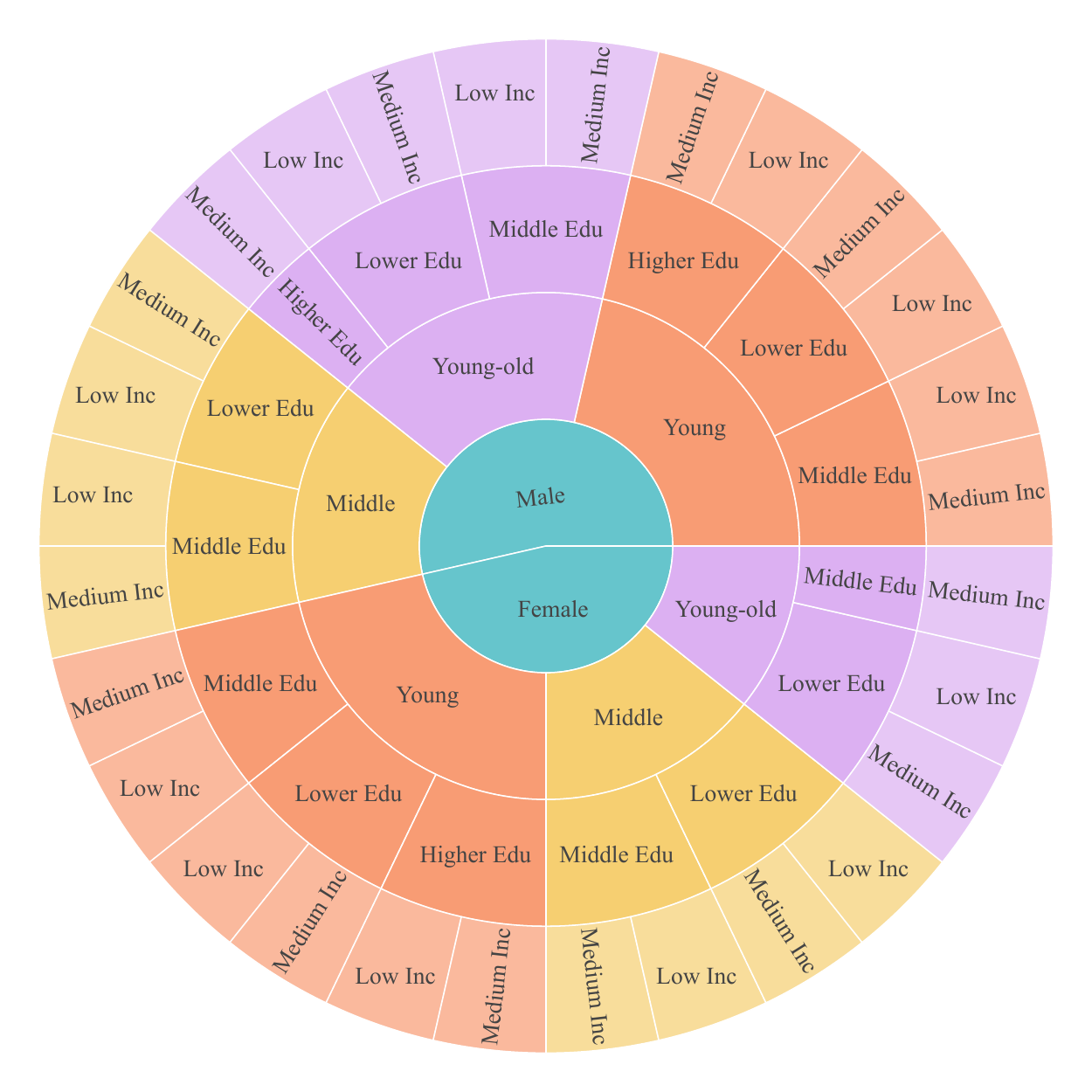}
    \caption{Final distribution of the 28 demographic groups in China, segmented by sex, age, highest educational level (Edu) and income level (Inc).}
    \label{fig:china_sunburst}
\end{figure}

\begin{figure}[ht]
    \centering
    \includegraphics[width=0.8\columnwidth]{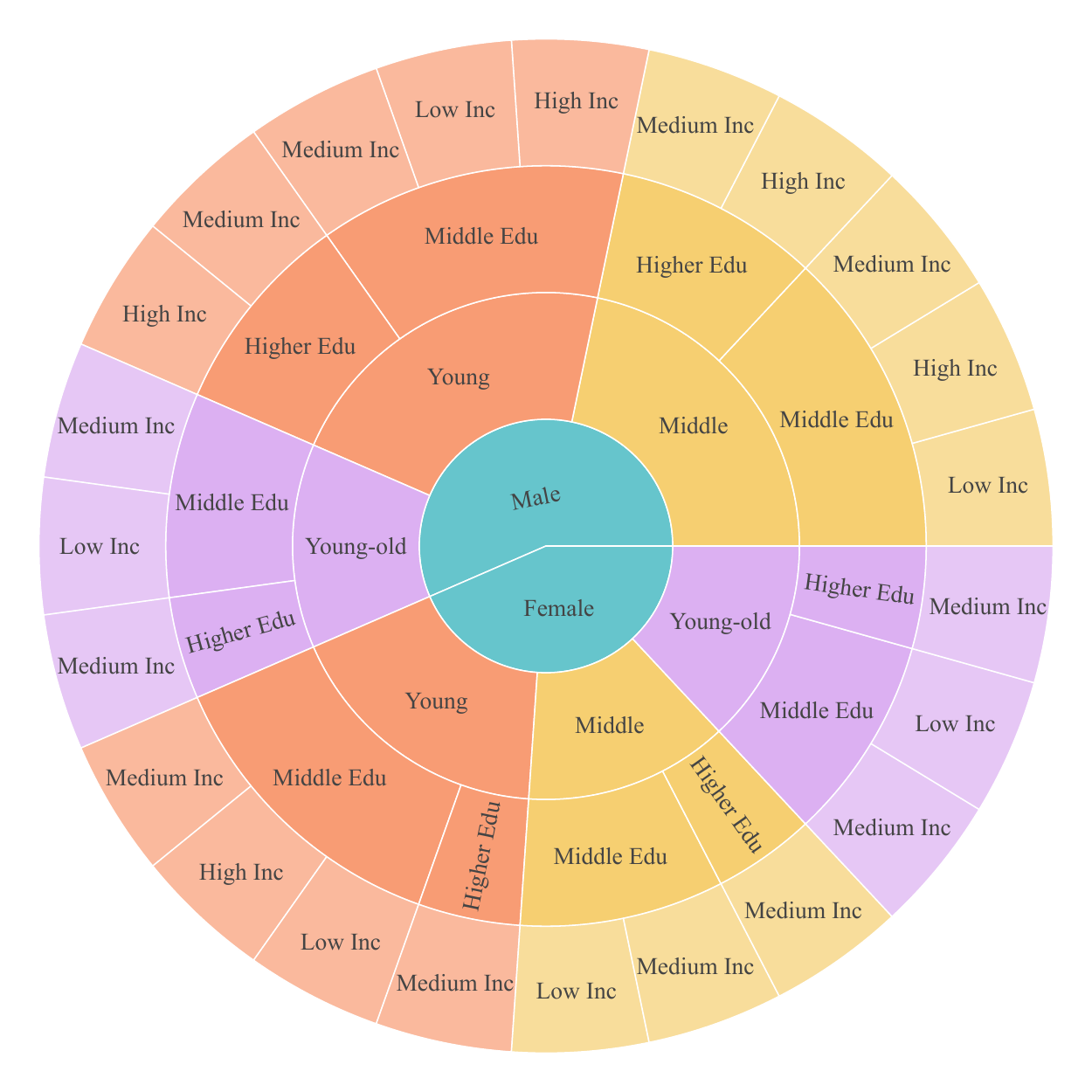}
    \caption{Final distribution of the 23 demographic groups in United States, segmented by sex, age, highest educational level (Edu) and income level (Inc).}
    \label{fig:us_sunburst}
\end{figure}

Figure \ref{fig:china_sunburst} and Figure \ref{fig:us_sunburst} illustrate the final group compositions for China and the United States, constructed based on four sociodemographic dimensions: age, gender, education level, and income level. This multi-dimensional grouping strategy enhances the discriminative power of group representations and provides a solid foundation for subsequent analyses of cross-group value differences and temporal dynamics.

Table \ref{tb:demographic_stats_china} and Table \ref{demographic_stats} present the sample sizes of the 28 Chinese groups and 23 U.S. groups across waves 5, 6, and 7 of the World Values Survey. To ensure the statistical reliability and representativeness of group-level analysis, we carefully examine the sample distributions across all waves. Groups with insufficient sample sizes in one or more waves are considered unreliable proxies for the corresponding demographic populations and are therefore excluded from subsequent experiments. This filtering process helps reduce noise introduced by sparse data and improves the robustness and stability of our simulation results.

\begin{table}[ht]
\centering
\caption{\textbf{Demographic Group Distribution across Waves (China).} 
This table compares the sample sizes of representative demographic groups in Wave~7, Wave~6, and Wave~5.}
\label{tb:demographic_stats_china}
\resizebox{0.48\textwidth}{!}{
\begin{tabular}{@{}cllllrrr@{}}
\toprule
\textbf{ID} & \textbf{Gender} & \textbf{Age} & \textbf{Edu.} & \textbf{Income} & \textbf{W7} & \textbf{W6} & \textbf{W5} \\ \midrule
    C1  & Female & Middle Age & Lower  & Low    & 173 & 56  & 88  \\
    C2  & Female & Middle Age & Lower  & Medium & 194 & 68  & 82  \\
    C3  & Female & Middle Age & Middle & Low    & 32  & 45  & 29  \\
    C4  & Female & Middle Age & Middle & Medium & 57  & 88  & 38  \\
    C5  & Female & Young-Old  & Lower  & Low    & 130 & 62  & 53  \\
    C6  & Female & Young-Old  & Lower  & Medium & 133 & 80  & 50  \\
    C7  & Female & Young-Old  & Middle & Medium & 26  & 26  & 14  \\
    C8  & Female & Young      & Higher & Low    & 50  & 35  & 6   \\
    C9  & Female & Young      & Higher & Medium & 228 & 113 & 40  \\
    C10 & Female & Young      & Lower  & Low    & 132 & 33  & 107 \\
    C11 & Female & Young      & Lower  & Medium & 172 & 40  & 96  \\
    C12 & Female & Young      & Middle & Low    & 51  & 102 & 71  \\
    C13 & Female & Young      & Middle & Medium & 158 & 217 & 128 \\
    C14 & Male   & Middle Age & Lower  & Low    & 122 & 41  & 73  \\
    C15 & Male   & Middle Age & Lower  & Medium & 135 & 59  & 61  \\
    C16 & Male   & Middle Age & Middle & Low    & 47  & 63  & 47  \\
    C17 & Male   & Middle Age & Middle & Medium & 59  & 82  & 68  \\
    C18 & Male   & Young-Old  & Higher & Medium & 13  & 9   & 6   \\
    C19 & Male   & Young-Old  & Lower  & Low    & 98  & 43  & 53  \\
    C20 & Male   & Young-Old  & Lower  & Medium & 107 & 65  & 24  \\
    C21 & Male   & Young-Old  & Middle & Low    & 22  & 20  & 17  \\
    C22 & Male   & Young-Old  & Middle & Medium & 32  & 47  & 23  \\
    C23 & Male   & Young      & Higher & Low    & 55  & 16  & 7   \\
    C24 & Male   & Young      & Higher & Medium & 176 & 106 & 37  \\
    C25 & Male   & Young      & Lower  & Low    & 96  & 37  & 61  \\
    C26 & Male   & Young      & Lower  & Medium & 103 & 28  & 33  \\
    C27 & Male   & Young      & Middle & Low    & 48  & 101 & 66  \\
    C28 & Male   & Young      & Middle & Medium & 129 & 217 & 140 \\ 
    \bottomrule
\end{tabular}
}
\end{table}

\clearpage

\begin{table}[ht]
\centering
\caption{\textbf{Demographic Group Distribution across Waves (U.S.).} This table compares the sample sizes of representative demographic groups in Wave 7, Wave 6 and Wave 5.}
\label{demographic_stats}
\resizebox{0.48\textwidth}{!}{
\begin{tabular}{@{}cllllrrr@{}}
\toprule
\textbf{ID} & \textbf{Gender} & \textbf{Age} & \textbf{Edu.} & \textbf{Income} & \textbf{W7} & \textbf{W6} & \textbf{W5} \\ \midrule
    U1  & Female & Middle Age & Higher & Medium & 68 & 159 & 13 \\
    U2  & Female & Middle Age & Middle & Low & 52 & 43 & 26 \\
    U3  & Female & Middle Age & Middle & Medium & 90 & 83 & 105 \\
    U4  & Female & Young-Old & Higher & Medium & 24 & 95 & 7 \\
    U5  & Female & Young-Old & Middle & Low & 29 & 35 & 23 \\
    U6  & Female & Young-Old & Middle & Medium & 71 & 91 & 69 \\
    U7  & Female & Young & Higher & Medium & 304 & 219 & 20 \\
    U8  & Female & Young & Middle & High & 18 & 7 & 11 \\
    U9  & Female & Young & Middle & Low & 101 & 45 & 53 \\
    U10 & Female & Young & Middle & Medium & 250 & 84 & 168 \\
    U11 & Male & Middle Age & Higher & High & 29 & 23 & 7 \\
    U12 & Male & Middle Age & Higher & Medium & 149 & 132 & 18 \\
    U13 & Male & Middle Age & Middle & High & 9 & 8 & 9 \\
    U14 & Male & Middle Age & Middle & Low & 39 & 44 & 28 \\
    U15 & Male & Middle Age & Middle & Medium & 111 & 81 & 110 \\
    U16 & Male & Young-Old & Higher & Medium & 132 & 116 & 7 \\
    U17 & Male & Young-Old & Middle & Low & 29 & 23 & 17 \\
    U18 & Male & Young-Old & Middle & Medium & 82 & 59 & 64 \\
    U19 & Male & Young & Higher & High & 45 & 28 & 7 \\
    U20 & Male & Young & Higher & Medium & 286 & 182 & 17 \\
    U21 & Male & Young & Middle & High & 14 & 18 & 11 \\
    U22 & Male & Young & Middle & Low & 65 & 49 & 49 \\
    U23 & Male & Young & Middle & Medium & 159 & 104 & 159 \\
    \bottomrule
    \end{tabular}
    }
\end{table}

\section{Event Impact Analysis Prompt Template} \label{event_impact}
\begin{figure}[h]
    \centering
    \includegraphics[width=1\columnwidth]{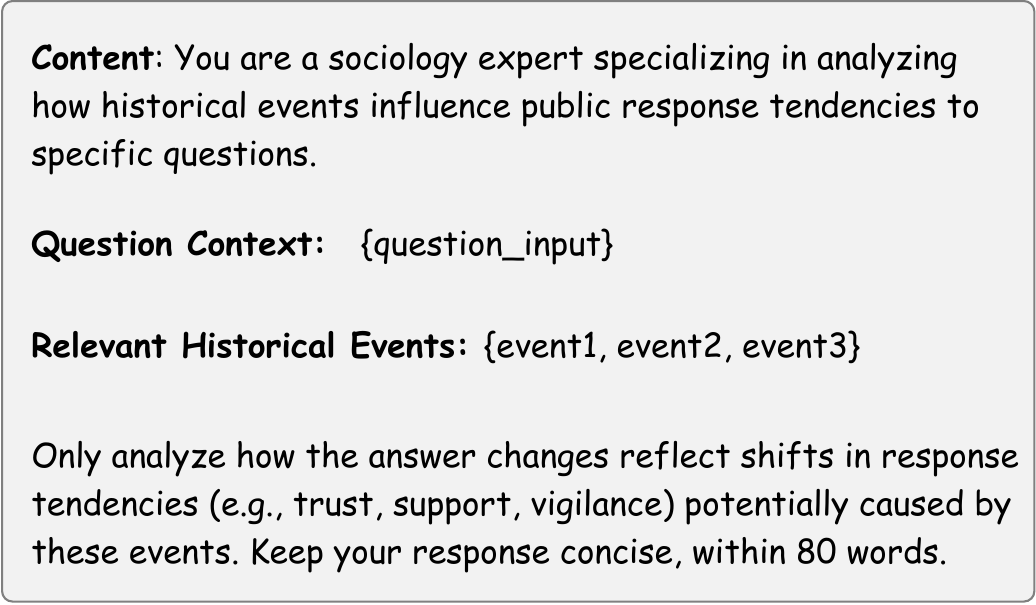}
    \caption{Event Impact Analysis Prompt Template.}
    \label{impact1}
\end{figure}

Figure \ref{impact1} illustrates the construction of the prompt template for event impact analysis. For each candidate event, we leverage large language models to infer its potential influence on different groups regarding specific issues. The quantified impact is encoded into a structured prompt and combined with historical value trajectories, enabling the model to jointly consider event-driven shocks and intrinsic group tendencies, thereby producing more realistic simulations of group-level value dynamics.

\section{Example of Chinese and English Prompts} \label{prompt}
\begin{figure}[ht]
    \centering
    \includegraphics[width=1\columnwidth]{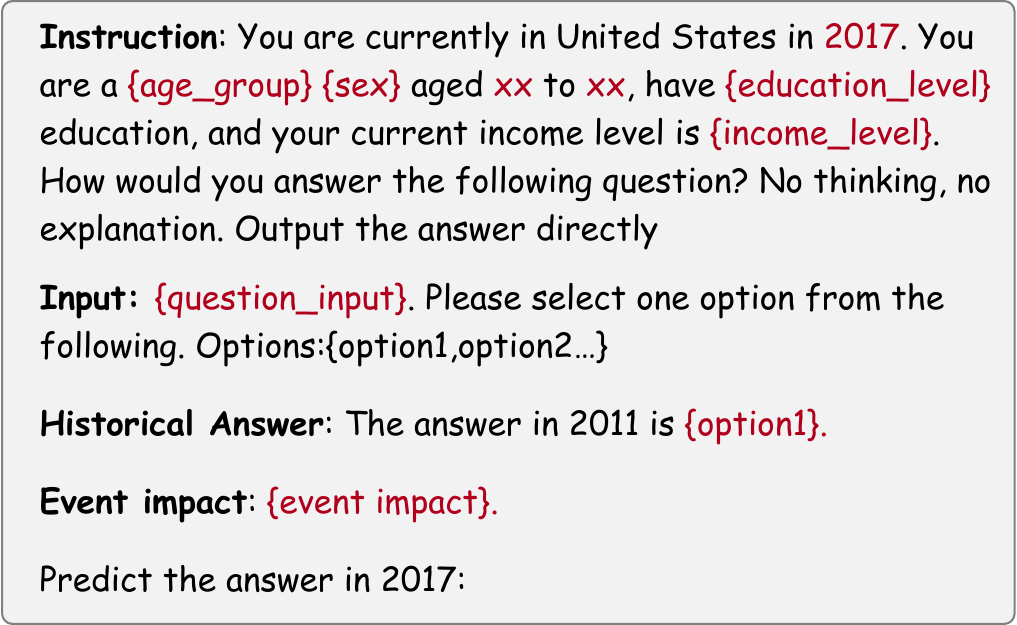}
    \caption{Example of an English prompt. The input variables are highlighted in red.}
    \label{fig:prompt_english}
\end{figure}

\begin{figure}[ht]
    \centering
    \includegraphics[width=1\columnwidth]{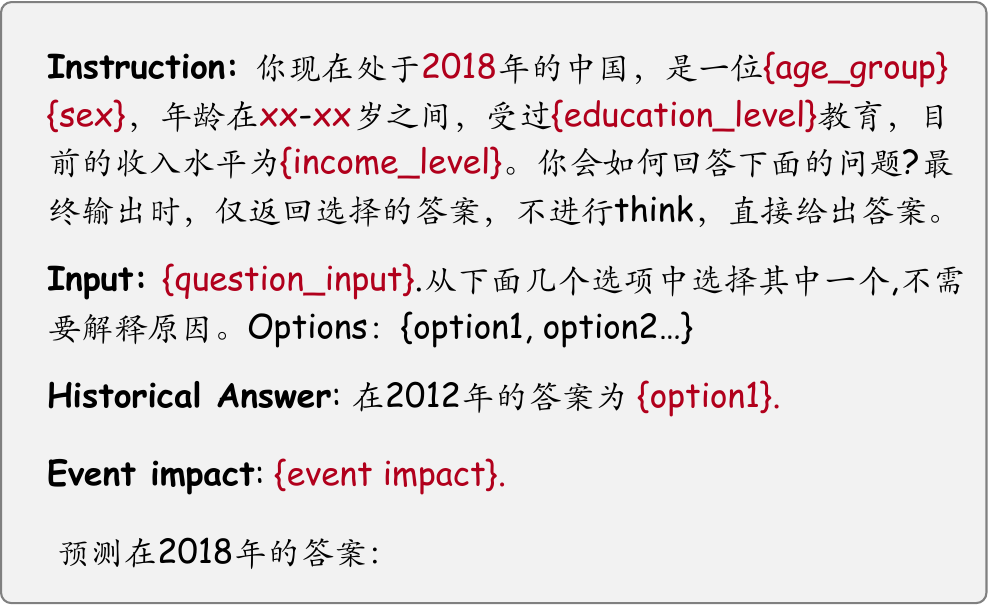}
    \caption{Example of a Chinese prompt. The input variables are highlighted in red.}
    \label{fig:prompt_chinese}
\end{figure}

Figures \ref{fig:prompt_english} and \ref{fig:prompt_chinese} illustrate the prompt templates used for social simulation in Chinese and U.S. cultural contexts, respectively.

\section{Models and Inference Settings} \label{setup}
Model Download we list all used open-source models here:


\definecolor{linkblue}{rgb}{0, 0, 0.55}

\begin{itemize}
    \item \textbf{Qwen3-8B}: \href{https://huggingface.co/Qwen/Qwen3-8B}{\texttt{\textcolor{linkblue}{https://huggingface.co/Qwen/Qwen3-8B}}}
    
    \item \textbf{Qwen3-14B}: \href{https://huggingface.co/Qwen/Qwen3-14B} {\texttt{\textcolor{linkblue}{https://huggingface.co/Qwen/Qwen3-14B}}} 
    
    \item \textbf{GLM4-9B-0414}: \href{https://huggingface.co/zai-org/GLM-4-9B-0414}
    {\textcolor{linkblue}{\nolinkurl{https://huggingface.co/zai-org/GLM-4-9B-0414}}}

    \item \textbf{Llama3.1-8B-Instruct}: \href{https://huggingface.co/meta-llama/Llama-3.1-8B-Instruct}
    {\textcolor{linkblue}{\nolinkurl{https://huggingface.co/meta-llama/Llama-3.1-8B-Instruct}}}

    \item \textbf{Mistral3-7B-Instruct-V0.3}: \href{https://huggingface.co/mistralai/Mistral-7B-Instruct-v0.3}
    {\textcolor{linkblue}{\nolinkurl{https://huggingface.co/mistralai/Mistral-7B-Instruct-v0.3}}}

\end{itemize}

In all steps, the top-p and temperature parameters are selected within the range of (0,1). All experiments are conducted on a Tesla A40 GPU with a single card of 48G video memory. We employ LoRA with a rank of 8, a scaling factor lora\_alpha set to 32, and a dropout rate of 0.08.

\section{Major Event Impact Case Studies} \label{case}

\begin{figure}[ht]
    \centering
    \includegraphics[width=1\columnwidth]{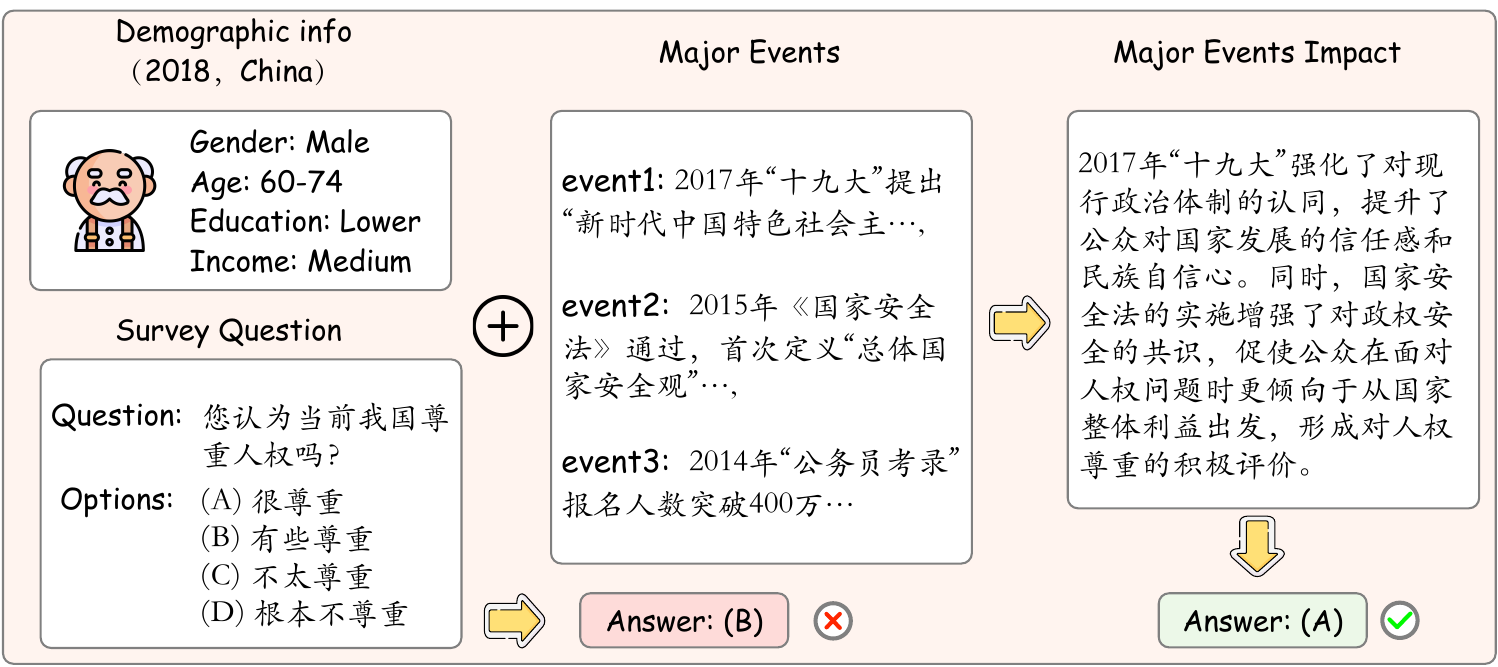}
    \caption{Case Study of the Impact of Major Events in China.}
    \label{fig:case1}
\end{figure}

\begin{figure}[ht]
    \centering
    \includegraphics[width=1\columnwidth]{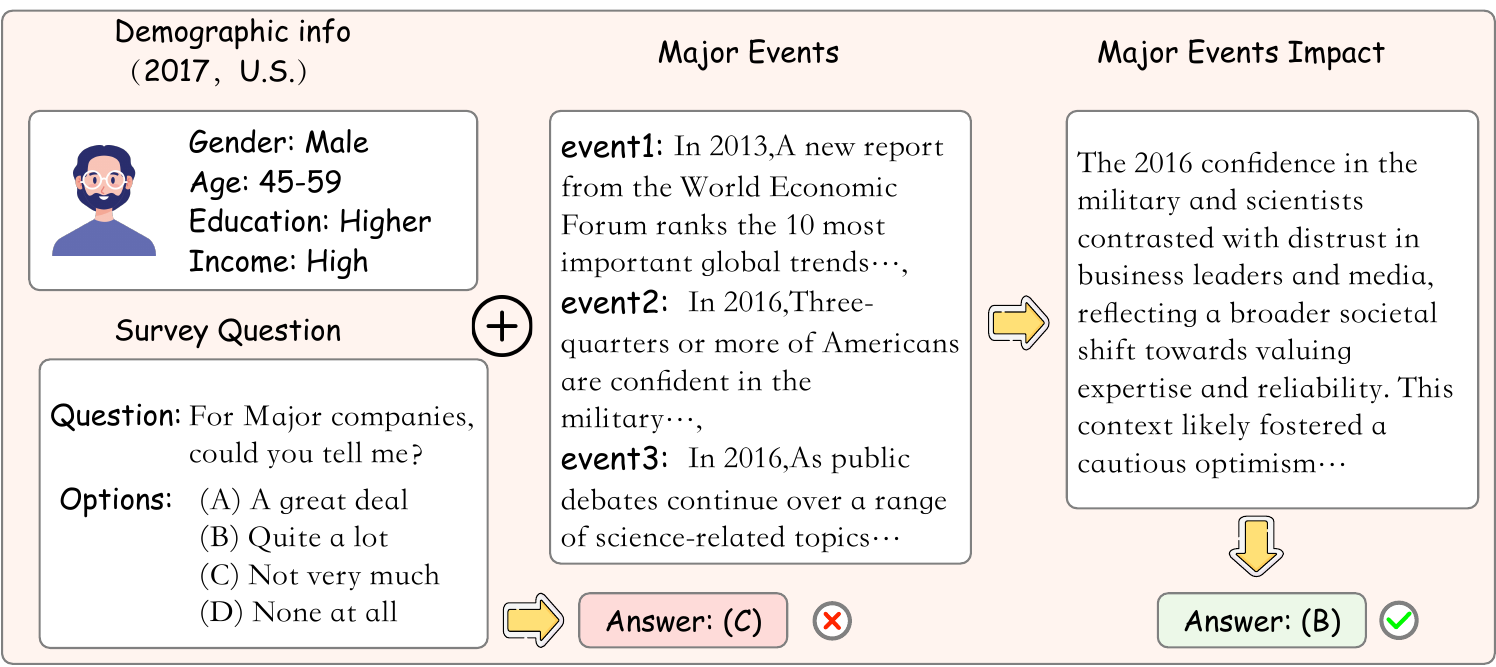}
    \caption{Case Study of the Impact of Major Events in United States.}
    \label{fig:case2}
\end{figure}

Figures \ref{fig:case1} and \ref{fig:case2} show how major events affect value responses among Chinese social groups, illustrating the role of event-driven dynamics in group value simulations.

\end{document}